\documentclass[aps,prd,showpacs,superscriptaddress,groupedaddress]{revtex4}

\usepackage{bm,graphicx}
\usepackage{amssymb,amsmath}   
\usepackage[utf8]{inputenc}
\usepackage{hyperref}

\renewcommand{\d}{\partial}

\newcommand{\nn}{\nonumber\\}

\newcommand{\exv}[1]{\left\langle{#1}\right\rangle}

\newcommand{\ep}{\varepsilon}

\newcommand{\p}{{\bf p}}
\newcommand{\T}{\textrm{T}}
\newcommand{\Tr}{\mathop{\textrm{Tr}}}

\newcommand{\pint}[2]{{\int\!\frac{d^{#1}#2}{(2\pi)^#1}\,}}

\newlength{\szovszel}
\newlength{\slashszel} 
\newcommand*{\sls}[1]{\mbox{%
    \settowidth{\szovszel}{\ensuremath{#1}}%
    \settowidth{\slashszel}{\ensuremath{\slash}}%
    \hspace*{0.5\szovszel}%
    \hspace*{-0.5\slashszel}%
    \slash%
    \hspace*{-0.5\szovszel}%
    \hspace*{-0.5\slashszel}%
    \ensuremath{#1}%
  }}

\newcommand{\tehat}{{\ensuremath{\quad\Rightarrow\quad}}}

\newcommand{\G}{{\cal G}}
\newcommand{\M}{{\cal M}}
\newcommand{\K}{{\cal K}}

\begin{document}

\title{Bound states in Functional Renormalization Group}

\author{Antal Jakov\'ac}
\email{jakovac@caesar.elte.hu}
\affiliation{Institute of Physics, E\"otv\"os Lor\'and University, 1/A P\'azm\'any P. s\'et\'any, H-1117 Budapest, Hungary}
\author{Andr\'as Patk\'os}
\email{patkos@galaxy.elte.hu}
\affiliation{Institute of Physics, E\"otv\"os Lor\'and University, 1/A P\'azm\'any P. s\'et\'any, H-1117 Budapest, Hungary}

\date{\today}

\begin{abstract}
Equivalence criteria are established for an effective Yukawa-type theory of composite fields representing two-particle fermion bound states with the original "microscopic" theory of interacting fermions based on the spectral decomposition of the 2-to-2 fermion scattering amplitude. Functional renormalisation group equations of the effective theory are derived exploiting relations expressing the equivalence. The effect of truncating the spectral decomposition is investigated quantitatively on the example of the non-relativistic bound states of two oppositely charged fermi particles. 
\end{abstract}


\maketitle

\section{Introduction}

Bound state formation in field theories is a fundamental problem. This is particularly valid in case of strong interactions, where the set of low energy observables is restricted exclusively to bound states of the particles (quarks and gluons) defining the theory.

The nonrelativistic approach to the bound state formation, i.e. the solution of the Schr\"odinger-equation, works nicely for atomic physics, also in heavy quark -- antiquark systems, but for relativistic systems it can not be generalized directly. The main reason is that, because of the retardation of the potential, the Lagrangian becomes nonlocal in time, and this makes the definition of the Hamiltonian cumbersome. Moreover, the propagators of the constituents are not restricted to the mass shell rigidly, and this modifies the naive potential (loop diagrams and ``crossed leg'' diagrams). As a consequence we are faced with a 2-particle, time-nonlocal problem with an improved potential known as the \emph{Bethe-Salpeter equation} (BSE) \cite{Salpeter:1951sz,Itzykson:1980rh}. This method was successfully used in the context of many QCD-related questions, c.f. for example \cite{Maris:1997piK,Maris:1999vmd,Hilger:2014nma}.

The generalized potential approach of the BSE is still not fully consistent, since the crossing symmetry of the relativistic quantum field theories is not obeyed. Technically speaking, in the diagrammatic expansion only the s-channel ladder diagrams are summed up, the t- and u-channel exchange (if there is allowed such) is treated only perturbatively. 

Because of these problems one employs also other frameworks to treat the bound state problem. One such framework is the infinite set of Dyson-Schwinger equations (DSE) \cite{Alkofer:2000wg,Sanchis-Alepuz:2015tha}. The ladder-type structures of the BSE resummation appear in the rainbow-ladder approximation of the DSE \cite{Eichmann:2016yit,Hilger:2017jti}. Carrying this approximation consistently for the 4-point function should provide the appropriate bound states of the system. Attempts to step beyond the ladder-summation are based on $nPI$ equations, where also the dynamical evolution of the interaction vertices is included \cite{Watson:2004kd,sanchis-alepuz:2015plb,williams:2016prd}.

The Functional Renormalization Group (FRG) equations \cite{Wetterich:1992yh,Morris:1993qb}, which are in principle exact (for reviews see \cite{gies:2012lnp,braun:2012jpg}), must also account for bound state formation. The most common practice is to choose an 
Ansatz for the effective quantum action accounts for both the fundamental and the bound state degrees of freedom and avoids double counting. Seminal works by Ellwanger and Wetterich \cite{Ellwanger:1993mw,Ellwanger:1994wy} have shown that a momentum-dependent 4-particle FRG equation implies the BSE. An efficient algorithm was also proposed and solved numerically for these equations. A disadvantage of this method is that the nonlocal 4-point function is a very complicated object of 
6 variables. A simplified approximation scheme preserving the momentum dependence of the three- and four-point functions has been proposed \cite{Blaizot:2005xy,Benitez:2011xx,Blaizot:2005wd,Blaizot:2006vr}. It has been applied to check the existence of bound states in the broken symmetry phase of the $\varphi^4$ model in three dimensions \cite{rose16}, proposed earlier on the basis quantum mechanics and BSE considerations \cite{Caselle:2000yx,Caselle:2001im}.
 
A general alternative approach characterizes the effective action with local terms which include also the interaction with the would-be bound states \cite{Jungnickel:1995fp, Gies:2001nw}. One can maintain this extended expression of the action during the whole scale evolution applying Hubbard-Stratonovich transformation after each FRG step to keep only the representatives of the bound states. This method, known also as dynamical hadronization, was used in several QCD studies \cite{Pawlowski:2005xe, Floerchinger:2009uf, Floerchinger:2010da, Alkofer:2018guy}. The choice of the fields and their masses extracted from two-point functions correspond to the phenomenological hadron data. To our knowledge it was not checked if the same pole singularities with the same masses appear in the 4-quark functions.

Functional Renormalisation Group techniques were applied also to investigating bound state signatures in the spectral representation of specific 2-point functions. Solutions of the flow equations of these functions \cite{Floerchinger:2009uf, Floerchinger:2010da, Kamikado2014} were continued analytically to Minkowski metrics where the imprints of the bound state poles were directly searched for. A common feature of these investigations is the truncation of the effective action at quartic level. Then one arrives at a closed system of flow equations for $\Gamma^{(2)}$ by imposing specific assumptions on the momentum dependence of $\Gamma^{(3)}$ and $\Gamma^{(4)}$ as they appear in those equations. In this way satisfactory mesonic spectral functions were determined in quark-meson models \cite{tripolt14,wambach14} but their consistency with the s-channel analytic structure of the 4-quark vertices has not been checked. Similar lack of the consistency check characterizes the FRG-reconstruction of the bound state found in the broken symmetry phase of the three-dimensional $\varphi^4$-theory \cite{rose16}.

In this paper we propose a more systematic approach to the introduction of the composite (bound state) fields based on the spectral representation of the 4-point function. It is shown in section II, that the most natural is to associate with each spectral eigencomponent of $\Gamma^{(4)}$ a composite field, therefore for an exact treatment we need infinite number of auxiliary fields. This observation is the central result of the present paper. In section III it is demonstrated that with a single composite field one cannot recover any pole indicating the presence of a bound state in the three-point coupling function of the composite field to its two constituents. In section IV the correct infinite set of FRG equations is presented (also in various approximations) for the two-point functions of the composite fields and the three-point functions connecting them to the two constituents.
 The masses emerging from the RG-flow in the theory defined with the composite fields are consistent by construction with the values one might find from the Bethe-Salpeter equation corresponding to the s-channel ladder resummation.  Also possible truncation of the RG equations is discussed. In section V the non-relativistic bound state problem of electrically oppositely charged fields is rewritten in form of an RG-flow equation. The accuracy of the lowest energy eigenvalue is investigated in function of the number of spectral components retained. The spirit of the present approach is close to the truncated conformal space approach \cite{Yurov:1989yu}, or its massive version \cite{Bajnok:2015bgw}.
The results of the paper are summarized in a concluding section (Section \ref{sec:concl}).
 
In order to make this paper nearly self-contained in an Appendix we shortly review the method of Bethe-Salpeter resummation and bring the non-relativistic bound state problem 
(Schr\"odinger-problem) to BSE form. An additional technical Appendix explains the solution of the non-relativistic BSE. A third Appendix is devoted to the operation of charge conjugation which is needed for the correct writing of the effective actions.

\section{Composite field representation of a fundamental field theory}
\label{sec:introbound}

Let us start our investigations with general considerations. Assume that we have some generic field theory, where there are (at least) two fermion species, and we want to examine the bound states of them. To be able to discuss these theories in general, concentrating on bound state formation, we \emph{integrate out exactly} all other degrees of freedom keeping the fermi fields intact, and get a pure two-fermion action with nonlocal vertices (cf. also \cite{rose16}). In case of QED, our main example below, the integration of the photon field is very simple, it only introduces a nonlocal 4-fermion term into the UV action. In case of QCD the integration of the gluons, because of the autonomous gluon dynamics, leads of course to much more complicated higher $n$-fermion couplings.

Now let us assume that we also have performed the path integral with the aforementioned pure fermionic theory, and obtained the 1PI effective action, which is a functional of the fermionic fields. For the moment it is not too important, how we get this effective action, this we will discuss in later sections, now we just assume that we know, at least in some approximation, this IR action.

In order to discuss 2-particle bound states, it is enough to truncate the action at the 4-fermion level, and treat
\begin{equation}
  \label{eq:Gamma}
  \Gamma_{eff} = \int_p\left[ \psi^\dagger_p {\cal K}^{(\psi)}_{p} \psi_p + 
    \chi^\dagger_p {\cal K}^{(\chi)}_{p} \chi_p\right] +
  \pint4p\frac{d^4q}{(2\pi)^4}\frac{d^4\ell}{(2\pi)^4}
 \lambda^{\ell}_{p\alpha\gamma,q\beta\sigma}
  \psi^\dagger_{p\alpha} \psi_{q\beta}
  \chi^\dagger_{\ell-p,\gamma}\chi_{\ell-q,\sigma}.
\end{equation}
We used here Euclidean description as well the Fourier components of the two kinds of fermi fields $\psi$ and $\chi$. Fourier momenta $(p,q,l)$ appear formally as lower "vector-indices" together with the bispinor indeces denoted by Greek letters. Note that the set of independent terms can be reduced by taking into account the Fierz identities \cite{Jaeckel:2002rm}.

In principle this action contains all information about the possible bound states of the system. As discussed in the Introduction, bound state energy levels can be identified as poles in the 4-fermion propagator. With the notation of the Appendix \ref{sec:BSE} we should look for the poles of the 4-fermion amputed amplitude \eqref{eq:4fermamp}. As also discussed in the Appendix, in BS approximation the condition that this amplitude has a pole, leads to the BS equation \eqref{eq:BSE0}.

This method is adequate only to identify bound state energies. But usually our goal is more ambitious, we would like to use the bound states as effective field degrees of freedom, with a complete propagator, and with interactions with other composite fields including themselves. Therefore we should propose an effective field theory containing also new bosonic bound state degrees of freedom that represent interacting two-particle states. We then require that the new effective theory should represent the same physics as the one with the original degrees of freedom. ``Same physics'' here means identical $n$-fermion connected correlation functions.

To accomplish this task we should use an effective model with auxiliary fields $H_{pn}$, where $p$ is its momentum, $n$ counts the different fields. The size of the set of auxiliary fields will be specified below. The auxiliary fields are coupled to the two constituent fermions through a Yukawa-type interaction. The Ansatz reads
\begin{equation}
  \label{eq:Gammabar}
  \Gamma_{eff} = \int_p\left[ \psi^\dagger_p {\cal K}^{(\psi)}_{p} \psi_p + 
    \chi^\dagger_p {\cal K}^{(\chi)}_{p} \chi_p+  \sum_{n}
    H^\dagger_{p n} \K^{(H)}_{p n} H_{p n}\right] +\sum_{n}
  \int_{p\ell}\left[ 
    H_{\ell n}  \psi^\dagger_{p\alpha} v^{\ell n}_{p \alpha\beta}
    \chi^*_{\ell-p,\beta} + H^\dagger_{\ell n} \chi^T_{\ell-p,\alpha}
    v^{\dagger\ell n}_{p \alpha\beta} \psi_{p,\beta}\right].
\end{equation}
 This Ansatz does not involve 4-fermion vertices. The interactions are mediated by the exchange of (perhaps infinitely many) auxiliary fields.
Our attention is restricted solely to $\psi-\chi$ pairs. 

The actual set of independent fermion--composite-field Yukawa-couplings $v$ is found by associating a different coupling with each irreducible Lorentz-representation built from the fermion bilinears.
We can observe that the combination $\psi^\dagger \bar \Gamma^{Rs} {\cal C}_E \chi^*$ is a vector operator in the representation of $R=\{S,P,V,A,T\}$ ($\Gamma^{Rs} = \{1, \gamma_5, \gamma_\mu, \gamma_5\gamma_\mu, \sigma_{\mu\nu}\}$ and $\bar \Gamma^{Rs}=\gamma_0 \Gamma^{Rs} \gamma_0$). ${\cal C}_E=\gamma_0\gamma_2$ is the charge conjugation operator
(cf. Appendix \ref{sec:App}). 

The Yukawa coupling also depends on the momentum in a generic way, so the p-integral in fact represents a form like
\begin{equation}
	\pint 4p f(p^2) p_{\mu_1}p_{\mu_2}\dots \bar \psi \Gamma^{Rs} C \bar\chi.
\end{equation}
This expression in general transforms as a product representation. Contracting the indices with properly chosen $T^{s,\mu_1,\mu_2,\dots}_n$ matrices we can split this form to a sum of irreducible representations, labeled by $n$. So in fact we have
\begin{equation}
   \pint 4p \psi^\dagger_{p\alpha} v^{\ell n}_{p \alpha\beta} \chi^*_{\ell-p,\beta} = T^{s,\mu_1,\mu_2,\dots}_n \pint 4p f(p^2) p_{\mu_1}p_{\mu_2}\dots \bar \psi \Gamma^{Rs} C \bar\chi.
\end{equation}
In this way the fermionic part is in an irreducible representation which can be combined with a multiplet $H_{\ell n}$ to form Lorentz scalars. 
Each of these Lorentz-invariant constructs has an independent coupling function $v$. 

Now that we specified the original 4-fermion theory as well as the one with new bosonic degrees of freedom, we can discuss the condition that they represent the same physics. Technically, we have to assure that the \emph{connected, amputed 4-point functions} are the same in the two theories. Since $\Gamma_{eff}$ is an effective action, it contains the proper vertices, thus to find the connected correlation functions we have to take into account the 1PR diagrams. On the side of the pure fermion action \eqref{eq:Gamma}, the connected, amputed 4-point function is just the 4-fermion coupling $\lambda$. Thus the condition reads\footnote{We remark that we can arrive at the same form by solving the equation of motion for the $H$ fields:
  \[
    \K^{(H)}_{\ell n} H_{\ell n} + \chi^T_{\ell-p,\alpha}
    v^{\dagger\ell n}_{p \alpha\beta} \psi_{p,\beta} =0 \tehat 
    H_{n\ell} = - G^{H}_{n\ell}\chi^T_{\ell-p,\alpha}
    v^{\dagger\ell n}_{p \alpha\beta} \psi_{p,\beta}.
    \]
    By substituting it back into \eqref{eq:Gammabar} we find that we generate a 4-point function similar to the one present in \eqref{eq:Gamma}.}
\begin{equation}
  \label{eq:reprcond}
  \lambda^{\ell}_{p\alpha\gamma,q\beta\sigma} = -\sum_n v^{\ell
    n}_{p\alpha\gamma} G^{(H)}_{\ell n} v^{\dagger\ell n}_{q\sigma\beta},
\end{equation}
or, in matrix notation:
\begin{equation}
  \label{eq:lambdarep}
  \bm{\lambda}^l = - \sum_n G^{(H)}_{ln} \bm{v}^l_n\otimes \bm{v}^{l\dagger}_n.
\end{equation}
In the index $\ell$ the matrix is block-diagonal, therefore we omit it to simplify the notation.
 
The most simple solution to fulfill this requirement is based on the spectral representation of $\bm{\lambda}$. Indeed, if we may choose $\bm v$ to be the \emph{eigenvectors} of $\bm\lambda$ with eigenvalues denoted by $G^{(H)}_n$
\begin{equation}
  \label{eq:eigensystem}
  \bm{\lambda}\bm v_n = -G^{(H)}_n \bm v_n,\qquad |\bm v_n|^2=1.
\end{equation}
then the spectral representation automatically leads to \eqref{eq:lambdarep}. We note here that it is a complete representation, which means in the language of quantum mechanics that both the bound states and the scattering states show up in the above sum. This means that some of the $H$ fields represent the bound states, but some of them represent only the common propagation of a loosely connected fermion pair. 

This equation can also be interpreted as the relativistic generalization of the Schr\"odinger-equation. In the nonrelativistic limit the time independent Schr\"odinger equation reads
\begin{equation}
  \bm H \psi_n = E_n\psi_n,\qquad |\psi_n|^2=1,
\end{equation}
where, in Fourier representation, $\psi$ depends on the 3-momentum $\bm p$. Formally therefore the Schr\"odinger equation is the same as equation \eqref{eq:eigensystem}, the role of wave function is played by the Yukawa coupling. The only difference is that now the wave function depends (besides the spinor indices) on the 4-momentum $p=(p_0,\p)$. Moreover, the eigenvalue is not the energy level, but the propagator of the auxiliary state, parametrized by the suppressed $\ell$ momentum.

This line of thought can be generalized and claim that the proper vertices of the pure fermion effective action can be interpreted as the ``Hamiltonians'' whose eigenvalues are the multiparticle composite state propagators. In particular, if we have a 6-fermion coupling with proper vertex $\Gamma^{(6)}$ (all indices suppressed), then it can be used as a ``Hamiltonian'' to determine the propagation of three-fermion composite operator. This composite operator can be represented by fermion fields with propagator $G^{(3)}$ and the coupling to the original fermionic degrees of freedom $v^{(3)}$, which therefore satisfy
\begin{equation}
  \bm\Gamma^{(6)} \bm v^{(3)} = -G^{(3)}\bm v^{(3)}.
\end{equation}
Therefore, in principle, we can write down the relativistic Schr\"odinger equation for any $n$-particle composite states (assuming "diagonal" propagation without any change in the particle type and number). These states are bound states, if their propagator contains poles.

As a closing remark we may observe that, although the discussed method is the most straightforward representation, we may generalize this setup. We can find a representation based not on the original $\bm{\lambda}$, but instead one makes use of its transformed form:
\begin{equation}
  \label{eq:eigen2}
  \bm A \bm{\lambda} \bm A^\dagger \bm x_n= c'_n \bm x_n,\qquad |
  \bm x_n|^2=1,
\end{equation}
where $\bm A$ is any (invertible) matrix. This provides a representation
\begin{equation}
  \bm A \bm{\lambda} \bm A^\dagger = \sum_n c'_n \bm x_n\otimes \bm
  x_n^\dagger,\qquad \bm{\lambda} = \sum_n c'_n(\bm A^{-1} \bm
  x_n)\otimes (\bm A^{-1} \bm x_n)^\dagger.
\end{equation}
This suggests that we can also choose a set of non-orthogonal $\bm{v}_n$ Yukawa-couplings:
\begin{equation}
  \label{eq:genrep}
  \bm{v}_n= \bm A^{-1} \bm x_n,\qquad G^{(H)}_n=-c'_n.
\end{equation}

Usually the eigenvalues also depend on the choice of $\bm A$ (unless it is unitary). Exceptions are the zero or infinite eigenvalues: for a zero eigenvalue that had originally an eigenvector $\bm x_0$, then
\begin{equation}
  \bm{\lambda}\bm x_0=0 \qquad\Rightarrow\qquad \bm A
  \bm{\lambda} \bm A^\dagger (\bm A^{-1\dagger} \bm x_0) =0,
\end{equation}
so  $\bm{\lambda} \bm A^\dagger$ has also a zero eigenvalue. Infinite eigenvalue strictly speaking means that the matrix does not exists, or that $\bm Q= \bm{\lambda}^{-1}$ is not invertible, since it has a zero eigenvalue. That also means that $\bm A^{-1\dagger} \bm Q \bm A^{-1}$ has also a zero eigenvalue, it is not invertible, so the corresponding inverse matrix $\bm A \bm{\lambda} \bm A^\dagger$ must have an ``infinite'' eigenvalue.

This means that although the auxiliary field propagators are not unique, their poles are representation independent, and so they can be considered as physical quantities. Therefore the bound state mass is well defined, but for example the bound state scattering amplitudes depend on the accurate definition of the off-shell parts of the bound
sates.

Summarizing the correspondence found in this section, we can faithfully
represent the 4-fermion interaction through a Yukawa-type theory with infinite number of auxiliary fields. The choice of these fields is not unique, but the poles of their propagators represent real physical singularities of the 4-point function. They provide unique characterization for what one should call the physical bound states.  The construction of the FRG equations for the bound state propagators and couplings is our final goal, which is discussed in the next two sections. 

\section{FRG-equation for a trial effective action of QED}
\label{sec:QEDint}

As an example we illustrate this process in case of QED. Since hermiticity will be important for what follows, we use an action with the conventional (not the Dirac) adjoint fields. The fundamental QED action of two, oppositely charged fermion species $\psi$ and $\chi$ read
\begin{equation}
  \label{eq:Sfund}
  S_{fund} = \int_p\left[ \psi^\dagger_p {\cal K}^{(\psi)}_{p} \psi_p + 
    \chi^\dagger_p {\cal K}^{(\chi)}_{p} \chi_p +
    \frac12 {A^a_p}^\dagger{\cal K}^{(A)}_p A^a_p +\int_q 
    (\psi^\dagger_p \gamma^a_{pq} \psi_q - \chi^\dagger_p\gamma^a_{pq} \chi_q)
    A^a_{p-q} \right],
\end{equation}
where
\begin{equation}
\label{eq:kernels}
  {\cal K}^{(\psi)}_{p} = \gamma_0(i\sls p +m_\psi),\qquad {\cal
    K}^{(\chi)}_{p} = \gamma_0(i\sls p+m_\chi),\qquad {\cal K}^{(A)}_p
  = p^2,\qquad \gamma^a_{pq} = -ie\gamma_0\gamma_E^a.
\end{equation}
Euclidean continuation is employed, where $a\in\{1,2,3,4\}$, $x_0=-ix_4$, $\d_0=i\d_4$, $A_0=iA_4$, $\gamma_j=i\gamma_{E}^j$ for $j<4$ and $\gamma_0=-\gamma_{E}^4$. Below a simplified notation is used, where the index 'E' is not displayed.
The inverse matrices of the kernels provide the propagators which will be denoted as $G^{(\psi)},\, G^{(\chi)}$ and $G^{(A)}$, respectively. In this way the propagators and the couplings (the photon-fermion vertices) all are hermitean matrices:
\begin{equation}
  G^{(\psi)\dagger}_{p\alpha\beta} = G^{(\psi)}_{p\alpha\beta},\qquad
  G^{(\chi)\dagger}_{p\alpha\beta} = G^{(\chi)}_{p\alpha\beta},\qquad
  \gamma^{a\dagger}_{p\alpha,q\beta} = 
  \gamma^{a*}_{q\beta,p\alpha} = \gamma^{a}_{p\alpha,q\beta}.
\end{equation}

\begin{figure}[htbp]
  \centering
  \includegraphics[height=1.8cm]{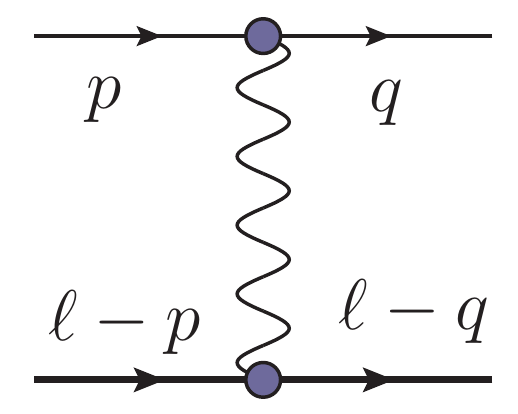}
  \caption{The tree level Feynman diagram contributing to the connected 4-point function. Thin lines represent $\psi$, thick lines $\chi$, and the curly line stands for the $A$-propagator.}
  \label{fig:fourpf_tree}
\end{figure}

The pure fermion theory \eqref{eq:Gamma} is matched at some scale $k=\Lambda$ with the original $\Gamma[\psi,\chi,A^a]$, which is the QED action \eqref{eq:Sfund} containing the electromagnetic coupling measured at that scale: $e(\Lambda)$. The matching condition for the connected $\psi$-$\chi$ 4-point function in the two models at momentum $\Lambda$ is based on simple one-photon exchange (cf. \eqref{eq:V0})
\begin{equation}
\label{eq:Mandlambda}
  V_{p\alpha\gamma,q\beta\sigma}^\ell =
  \lambda^{\ell}_{p\alpha\gamma,q\beta\sigma}\bigr|_{k=\Lambda},
\end{equation}
where the diagram contributing is given in Fig.~\ref{fig:fourpf_tree}. 
One notes that in the non-relativistic limit this exchange builds up the electrostatic potential between the two fermions.
The expression of $V^{\ell}_{p\alpha\gamma,q\beta\sigma}$ reads as
\begin{equation}
  \label{eq:V0}
  V^{\ell}_{p\alpha\gamma,q\beta\sigma} = \gamma^a_{p\alpha,q\beta}
  \gamma^a_{\ell-p\gamma,\ell-q\sigma} G^{(A)}_{p-q}.
\end{equation}


At this point we should stop for a while and think the whole strategy once again. If we have a 4-fermion term in the theory, then it is \emph{nonrenormalizable} by the canonical dimension counting, or, put another way, this operator is irrelevant in the IR. So we may ask, do we need this complicated term at all?

One can imagine a much simpler strategy. Since the bound state consists of a $\psi$ and a $\chi$ particle, we may just want to introduce one field degree of freedom for the ground state in a composite channel with specific spin and parity. With this action we take into account all relevant, low energy degrees of freedom of the system. The technical challenge is now to (dis)prove the existence of a bound state singularity in the singled-out channel?

The effective field, representing the ground state, will be called $H$. This bound state field couples to a $\psi$ and a $\chi$ particle to lowest order through a Yukawa-like coupling $\sim H \psi^\dagger\chi^\dagger$. In order to maintain relativistic invariance we have to include its charge conjugated counterpart, too. 

Then we can choose an FRG Ansatz that relies on the degrees of freedom $\psi,\,\chi,\,A^a$ and $H$, and contains relevant (renormalizable) operators. In Minkowski space it corresponds to the choice
\begin{equation}
  {\cal L} =- \frac14 F_{\mu\nu} F^{\mu\nu} + 
  \psi^\dagger\gamma_0(i\sls\nabla -m_\psi)\psi + 
  \chi^\dagger\gamma_0(i\sls\nabla^\dagger -m_\chi)\chi  +  H^\dagger
  \K^{(H)}(i\d) H - H  \psi^\dagger \tilde v \chi^* - H^\dagger \chi^T
  \tilde v^\dagger \psi,
\end{equation}
where the coupling function $\tilde v$ is a spin-dependent and
eventually nonlocal object, $\nabla$ stands for the covariant derivative. It actually depends also on the momentum of the bosonic bound state, but this is a spectator variable not indicated explicitly. To maintain Lorentz-invariance we may choose $\tilde v_{p\alpha\beta} = v_p{\cal C}_E$.  Changing over to the Euclidean theory in
Fourier space yields (suppressing spinor indices)
\begin{equation}
  \label{eq:Ansatz1}
  \Gamma = \Gamma_{QED} + H^\dagger_p \K^{(H)} H_p +
  H_\ell\psi_{p}^\dagger \tilde v^\ell_p \chi^*_{\ell-p} + H^\dagger_\ell
  \chi_{\ell-p}^T \tilde v_p^{\ell \dagger} \psi_{p}, 
\end{equation}
where the QED part is the same as in \eqref{eq:Sfund}, $K^{(H)}$ and
$v$ are the new scale-dependent parameters, characterizing the new functional pieces.

\begin{figure}[htbp]
  \centering
  \raisebox{0.55em}{\includegraphics[height=1.5cm]{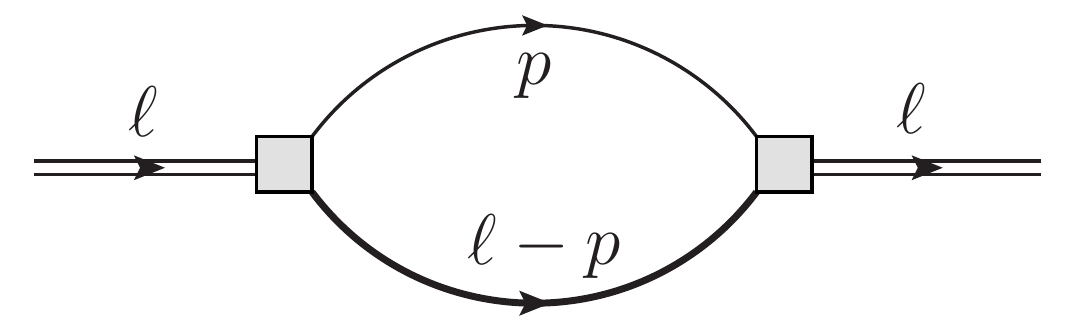}}
  \hspace*{2em}
  \includegraphics[height=2.2cm]{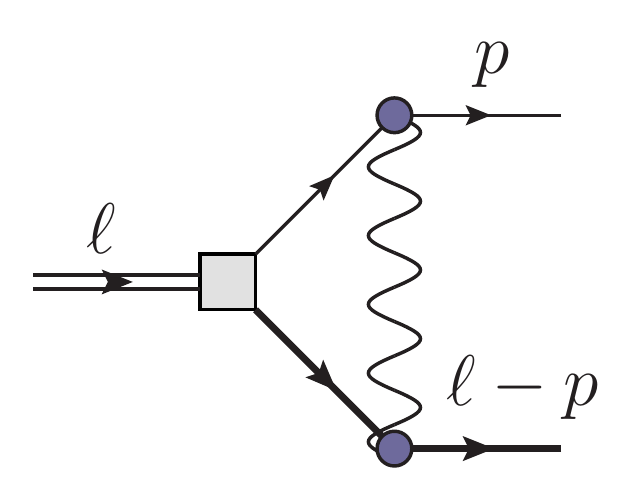}
  \caption{The one-loop corrections for the propagator and the vertex
    in the effective model. The box symbolizes the $v$ vertex.}
  \label{fig:tempt}
\end{figure}

A one-loop correction to the  self-energy of $H$ as well as to the
coupling $v$ can be computed by evaluating appropriate expectation
values. For the self-energy of $H$ we have to compute the amputed 2-point
function $-\exv{\hat H \hat H^\dagger}$ (c.f. Fig.~\ref{fig:tempt}).
The one-loop correction reads
\begin{equation}
  \Sigma_H(\ell)= - \pint4p \tilde v^{\ell*}_{p\alpha\beta} \tilde
  v^\ell_{p\alpha'\beta'} G^{(\psi)}_{p\alpha\alpha'}
  G^{(\chi)}_{\ell-p,\beta\beta'} = -\Tr \bm v^\dagger \bm \G \bm v, 
\end{equation}
where $\bm v$ is the vector notation of $\tilde v_{p\alpha\beta}^\ell$, the 2-fermion propagator is denoted by $\bm \G$. and
the trace as well as the adjoint is meant in the multi-index notation.

For the one loop correction of $v^\ell_p$ we consider the
following expectation value (c.f. Fig.~\ref{fig:tempt}):
\begin{equation}
  \exv{H'_\ell {\psi'}^\dagger_{p\alpha}{\chi'}^\dagger_{\ell-p,\beta}
  }_\mathrm{amputed}  = \tilde v_{p\alpha\beta}^\ell - \pint4p
  \gamma^a_{p\alpha,q\alpha'} G^{(\psi)}_{q,\alpha'\sigma}
  \tilde v^\ell_{q\sigma\sigma'} G^{(A)}_{p-q}
  \gamma^a_{\ell-p,\beta,\ell-q,\beta'} G^{(\chi)}_{\ell-q,\beta'\sigma'}. 
\end{equation}
 In matrix notation it can be written as
\begin{equation}
  \exv{H'_\ell {\psi'}^\dagger_{p\alpha}{\chi'}^\dagger_{\ell-p,\beta}
  }_\mathrm{amputed} = \bm v_{\alpha\beta} - (\bm V \bm\G)_{\alpha\beta,\sigma\sigma'}  \bm v_{\sigma\sigma'}.
\end{equation}
Note that with the above Ansatz there is no diagram that contains H-exchange in the t-channel, and so there is no $\sim v^3$ correction.

Now let us determine the FRG equations using this Ansatz. The general
Wetterich equation reads
\begin{equation}
  \d_k \Gamma = \frac12 \hat\d_k\Tr\ln (\Gamma^{(2)}+R_k),
\end{equation}
where $R_k$ is the regulator and the derivative with respect the scale,
 $\hat \d_k$ acts only on the regulator. Using this form one can easily
determine the equations for higher derivative $n$-point functions, since the corresponding one-loop diagrams just involve larger number of amputed external legs.
Therefore below we simply present the RG equations without entering into the derivation steps.

Since we want to concentrate on the bound states, 
only the evolution of the bound state kernel $\K^{(H)}$ and of the 2-fermion--bound state 3-point function
$\bm v$ is tracked, the fermion masses and the photon propagator are kept the same at all scales (for instance, the generated 4-fermion terms and the composite self-interaction are left out of consideration). We find:
\begin{eqnarray}
  && \d_k \K^{(H)}_p = - \hat\d_k \Tr\,\bm v^\dagger \bm\G \bm v \nn
  && \d_k \bm v =  -\hat\d_k(\bm V\bm\G) \bm v.
\end{eqnarray}
Now we can use the fact that the regulator affects only $\bm\G$ (the potential $V$ comes from the full photon exchange). Moreover, since we do not run the QED parameters, in this simple approximation $\hat\d_k\bm\G = \d_k\bm\G$ since there is no other $k$ dependence. So we obtain
\begin{eqnarray}
  \label{eq:temptFRG}
  && \d_k \K^{(H)}_\ell = -\Tr\,\bm v^\dagger \d_k \bm\G \bm v \nn
  && \d_k \bm v =  -\bm V (\d_k\bm\G) \bm v.
\end{eqnarray}
We remark that we would arrive at a similar expression in \emph{effective theories of strong nuclear interactions} where the photon-mediated 4-fermion interaction would have been substituted by a pion exchange.

The second equation is very similar to the derivative of the BS-equation \eqref{eq:BSE0}. The difference is that from the derivative with respect to $k$ of \eqref{eq:BSE0} we obtain
\begin{equation}
 \d_k \bm v =  -\bm V (\d_k\bm\G) \bm v - \bm V \bm\G \d_k\bm v  \tehat \d_k \bm v = -(1+\bm V \bm\G)^{-1} \bm V (\d_k\bm\G) \bm v.
\end{equation}

Thus we do not have the same equation as the one that would come from the BS-equation. And, correspondingly, we also do not have the same solution. The equation \eqref{eq:temptFRG} for $\bm v$, namely, can be solved symbolically in form of a "$k$-ordered" exponential: 
\begin{equation}
  \bm v_{k} =  \T_k e^{-\int_k^\Lambda dk' \bm V \bm\G(k')} \bm
  v_{k=\Lambda},\qquad\mathrm{vs.}\qquad \bm v_{BS,k} = (1+ \bm V\bm\G(k))^{-1}\bm v_{k=\Lambda}.
\end{equation}

As we see, the solution of the Ansatz of this section misses the pole in the fermion-bound state vertex which is the signature for the appearance of bound states. So, although tempting is the simplicity of Ansatz \eqref{eq:Ansatz1}, it is not adequate to account for the bound states.

To understand what is the conceptual problem with this approach we expand the exponential factor in powers of $\bm V\bm\G(k)$  and recognize the presence of the $1/n!$ suppression factor relative to the expansion of the BS-solution.

The origin of this suppression factor is the following. When we solve an equation $\d_k \bm x = \bm A(k) \bm x$ recursively, then we obtain a series of ladder diagrams. But the momentum of the subsequent ladder rungs is strictly ordered, since we can insert a rung only at the left end of the series containing rungs with higher $k$. Therefore the exponential function corresponds to ladder diagrams with rungs of ordered momenta. If we release the momentum ordering, then the rungs can appear in all possible sequence, yielding a factor of $n!$ growth at $n$th order.

\section{Functional Renormalisation Group evolution of the composite field action}

In section II we have presented the rules for the construction of a theory of composite fields in which bound states in a theory containing fundamental fermionic degrees of freedom show up explicitly. There we have argued that the 4-fermion interaction plays a crucial role in a correct construction, in which one can consistently access the bound states. In this section we develop the FRG method for the solution of the bound state theory \eqref{eq:Gammabar}.

A safe algorithm fullfilling the above requirement starts with the (temporary) integration over the composite boson fields. Then one performs the FRG-step in the purely fermi-representation. The RG-cycle is made complete by projecting back the result to the bosonic representation. This technique is very similar to the procedure of Gies and Wetterich \cite{Gies:2001nw}, Alkofer {\it et al.} \cite{Alkofer:2018guy} and  Braun {\it et al.} \cite{braun:2016dh}. The difference is that we perform the FRG-step in the pure fermion system, while in the quoted papers just the 4-fermion coupling is kept zero. Further difference is present in the non-trivial momentum dependence of the 4-point functions, and most importantly, the infinite set of auxiliary fields needed for a faithful representation of the 4-point function.

\subsection{FRG in a pure fermionic theory}
\label{sec:FRGpure}

In a preparatory step we work out the FRG evolution for the pure fermionic theory. Although in this case the bound state fields are not seen explicitly, but we may check the consistency of the FRG results against the BS resummation. This approach was followed by Refs. \cite{Ellwanger:1994wy, Ellwanger:1993mw, rose16}.

\begin{figure}[htbp]
  \centering
  \includegraphics[height=2.18cm]{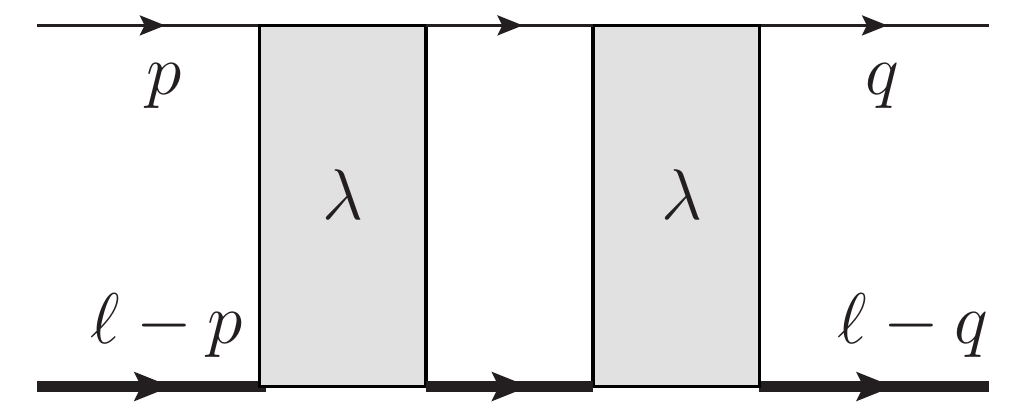}
  \hspace*{2em}
  \includegraphics[height=2.2cm]{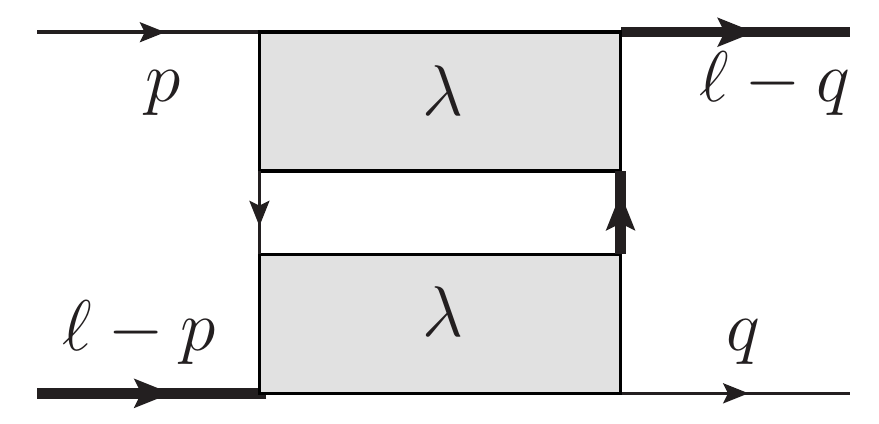}
  \caption{Vertex correction at one loop level in the 4-fermion effective theory. The box symbolizes the $\lambda$ vertex. The second diagram has the same topology as the perturbative crossed-leg diagram.}
  \label{fig:fermion}
\end{figure}

Now we turn to the general discussion (with no explicit reference to electrodynamics) and first compute radiative corrections to the four-point function.  We find for the 4-point 1PI function (c.f. Fig.~\ref{fig:fermion})
\begin{equation}
  \Gamma^{(4)\ell}_{p\alpha\gamma,q\beta\sigma} = 
  \lambda^\ell_{p\alpha\gamma,q\beta\sigma}-\pint4r\biggl[
  \lambda^\ell_{p\alpha\gamma,r\beta'\sigma'}
     \G^\ell_{r,\beta'\sigma'\alpha'\gamma'}
     \lambda^\ell_{r\alpha'\gamma',q\beta\sigma}
    +\lambda^{\ell+r-q}_{p\alpha\gamma',r\beta'\sigma}
     \G^{\ell+2k-p-q}_{r,\beta'\sigma'\alpha'\gamma'}
     \lambda^{\ell+r-p}_{r\alpha'\gamma,q\beta\sigma'} \biggr],
\end{equation}
where the correction has a very similar structure as $({\cal M}_{p\alpha\gamma,q\beta\sigma}^{\ell})^{1-loop}$ in \eqref{eq:Moneloop}. The first term in the square bracket corresponds to the ladder-type diagram, the second one corresponds to the crossed leg diagram.

For the FRG equation of the coupling $\lambda$ one finds:
\begin{equation}
  \d_k \lambda^\ell_{p\alpha\gamma,q\beta\sigma} = - \hat\d_k\pint4r\biggl[
   \lambda^\ell_{p\alpha\gamma,r\beta'\sigma'}
     \G^\ell_{r,\beta'\sigma'\alpha'\gamma'}
     \lambda^\ell_{r\alpha'\gamma',q\beta\sigma}
    +\lambda^{\ell+r-q}_{p\alpha\gamma',r\beta'\sigma}
     \G^{\ell+2k-p-q}_{r,\beta'\sigma'\alpha'\gamma'}
     \lambda^{\ell+r-p}_{r\alpha'\gamma,q\beta\sigma'} \biggr].
\end{equation}
Using that $k$-dependence comes solely from the regularization of $\G$, we can also write
\begin{equation}
  \label{eq:lambdabarFRG}
  \d_k \lambda^\ell_{p\alpha\gamma,q\beta\sigma} = -\pint4r\biggl[
   \lambda^\ell_{p\alpha\gamma,r\beta'\sigma'}
     \d_k \G^\ell_{r,\beta'\sigma'\alpha'\gamma'}
     \lambda^\ell_{r\alpha'\gamma',q\beta\sigma}
    +\lambda^{\ell+r-q}_{p\alpha\gamma',r\beta'\sigma}
     \d_k \G^{\ell+2r-p-q}_{r,\beta'\sigma'\alpha'\gamma'}
     \lambda^{\ell+r-p}_{r\alpha'\gamma,q\beta\sigma'} \biggr].
\end{equation}
The initial condition for the flow, as mentioned above, is $\lambda(k=\Lambda)=V$. This is the complete equation which, as it comes from a consistent Ansatz, respects all the symmetries of the quantum field theory like the crossing symmetry.

The Bethe-Salpeter (BS) resummation is known to violate crossing symmetry, and so we can not expect that the above result fully agrees with the BS results. Therefore to have a connection to the BS result, we have to simplify our FRG equations, taking into account only the first (ladder) term in the differential equation. Then the FRG equation can be written in matrix form as before:
\begin{equation}
  \label{eq:lambdabarFRG1}
  \d_k {\bm\lambda} = -{\bm\lambda}(\d_k\bm\G)
  {\bm\lambda},\qquad {\bm\lambda}(k=\Lambda)=\bm V. 
\end{equation}

This equation can be solved explicitly
\begin{equation}
  \d_k{\bm\lambda}^{-1} = \d_k\bm\G \tehat {\bm\lambda}_k^{-1}
  - {\bm\lambda}_\Lambda^{-1} = \bm\G_k - \bm\G_\Lambda.
\end{equation}
Assuming that $\bm\G(\Lambda)=0$ due to the regulator, we find
\begin{equation}
  \label{eq:lambdaFRG}
  {\bm\lambda}_k = (1+\bm V\bm\G_k)^{-1} \bm V.
\end{equation}
We see that at $k=0$ we obtain a pole in $\lambda_{k=0}$ if there is one eigenvector $u$ of $\bm V\bm\G_0$ with unit eigenvalue: $\bm V\bm\G_0u=-u$. Comparing to the Bethe-Salpeter equation reviewed in the Appendix, this condition coincides with the Bethe-Salpeter formulation of the bound state problem.  
This means that this reduced version of FRG contains enough information to reproduce the bound state spectrum at least in agreement with the Bethe-Salpeter approach. Moreover, equation (\ref{eq:lambdabarFRG}) is in addition a fully consistent resummed form.

\subsection{FRG equations of the bound state system: the BS   approximation}

In order to avoid technical complications, let us first discuss the RG-steps for the FRG equations that faithfully represent the BS approximation (s-channel approximation). We have seen earlier that eq. \eqref{eq:lambdabarFRG1} can reproduce the BS resummation of the Appendix. 
Therefore we will use this equation with the representation \eqref{eq:lambdarep}. Sandwiching the  evolution equation \eqref{eq:lambdabarFRG1} with the eigenvectors of $\bm{\lambda}$ we find 
\begin{equation}
  \bm v_m^\dagger (\d_k\bm{\lambda})\, \bm v_n =-\delta_{nm} \d_k G^{(H)}_n 
  -G^{(H)}_n \bm v_m^\dagger \d_k\bm v_n - G^{(H)}_m \d_k\bm v_m^\dagger  \bm v_n
  =- G^{(H)}_nG^{(H)}_m \bm v_m^\dagger(\d_k\bm\G)\bm v_n.
\end{equation}
Using the fact that $\bm v_n$ are orthonormal (cf. \eqref{eq:eigensystem}),
we have
\begin{equation}
  \delta_{nm} \d_k G^{(H)}_n + (G^{(H)}_n-G^{(H)}_m) \bm v_m^\dagger \d_k\bm v_n
  = G^{(H)}_nG^{(H)}_m \bm v_m^\dagger(\d_k\bm\G)\bm v_n.
\end{equation}
This yields
\begin{eqnarray}
  &&\d_k G^{(H)}_n= (G^{(H)}_n)^2 \bm v_n^\dagger(\d_k\bm\G)\bm v_n,\nn
  &&\d_k\bm v_n = \sum_{\ell\neq n}
     \frac{G^{(H)}_\ell G^{(H)}_n}{G^{(H)}_n-G^{(H)}_\ell} \bm v_\ell (\bm v_\ell^\dagger(\d_k\bm\G)\bm v_n).
\end{eqnarray}
It is convenient to work with the $H$-field kernels using $G^{(H)} = 1/\K^{(H)}$. We find
\begin{eqnarray}
  \label{eq:BSFRG}
   &&\d_k \K^{(H)}_n= -\bm v_n^\dagger(\d_k\bm\G\bm v_n,\nn
   &&\d_k\bm v_n = \sum_{m\neq n} \frac1{\K^{(H)}_m-\K^{(H)}_n} \bm v_m (\bm v_m^\dagger(\d_k\bm\G)\bm v_n).
\end{eqnarray}
Writing out the indices, the matrix elements of $\d_k\G$ read
\begin{equation}
  \bm v_m^\dagger(\d_k\bm\G)\bm v_n = v^{m\ell*}_{p\alpha\beta}
  (\d_k\G)^{\ell}_{p,\alpha\beta, \gamma\sigma} v^{n\ell}_{p\alpha\sigma}.
\end{equation}
To compute this expression we just need to perform a $p$ integral and spinor index summations. 

To provide an initial condition, we have to start the evolution with the 4-point function defined at the UV-cutoff. It might be chosen to result from the integration of a degree of freedom. Then this 4-point function is just the relativistic generalisation of the potential emerging from the $t$-channel exchange of the corresponding force fields. For example in QED the "potential" is $V$, defined in \eqref{eq:V0}, where the gamma matrices are momentum independent. Therefore for $V$ the eigenvalue equation is a convolution in the momentum space, or in other words, a product in the real space:
\begin{equation}
  \lambda(k=\Lambda)^{\ell}_{p\alpha\beta,q\gamma\sigma} =
  G^{(A)}_{p-q}\gamma^a_{\alpha\beta}\gamma^a_{\gamma\sigma} 
  \qquad\Rightarrow\qquad
  \lambda(k=\Lambda)^{\ell}_{x,\alpha\beta,\gamma\sigma} =
  G^{(A)}_{x}\gamma^a_{\alpha\beta}\gamma^a_{\gamma\sigma}.
\end{equation}
The eigenvalue equation can be solved by the Ansatz 
\begin{equation}
   v^n_{x\alpha\beta} = v^n_{xc} \gamma^c_{\alpha\beta}.
\end{equation}
Then we have
\begin{equation}
  4G^{(A)}_x v^n_{xc} = c_n v^n_{xc}\qquad\Rightarrow\qquad v^n_{xc} =  \delta(x-n),\qquad c_n = 4G^{(A)}_n. 
\end{equation}

The advantage of the representation with the bound/composite states lies in the fact that it may allow { to handle the information about the system much more economically}. The 4-point coupling has a momentum structure $\lambda^\ell_{pq}$. Taking into account Lorentz invariance this function may depend on $\ell^2,\, p^2,\, q^2,\, \ell p,\, \ell q$ and $pq$, i.e. six invariants. In the BS-approximation $\ell$ is a spectator index, so $\ell^2$ should not be taken into account, but even then we have a function with five arguments. The 3-point function, on the other hand, has a momentum and species structure $v^{n\ell}_p$, in the relativistic invariant case it may depend on $n,\, \ell^2,\, p^2$ and $\ell p$. These are four parameters. Further, in the BS approximation we have a function with only three arguments. Even better, one of these parameters is a species index, and if we trust the argument that IR physics is dominated by the eigenvectors with the smallest eigenvalues, then we may use a moderate number of species.

\subsection{The general FRG equations}

\begin{figure}[htbp]
  \centering
  \includegraphics[height=2cm]{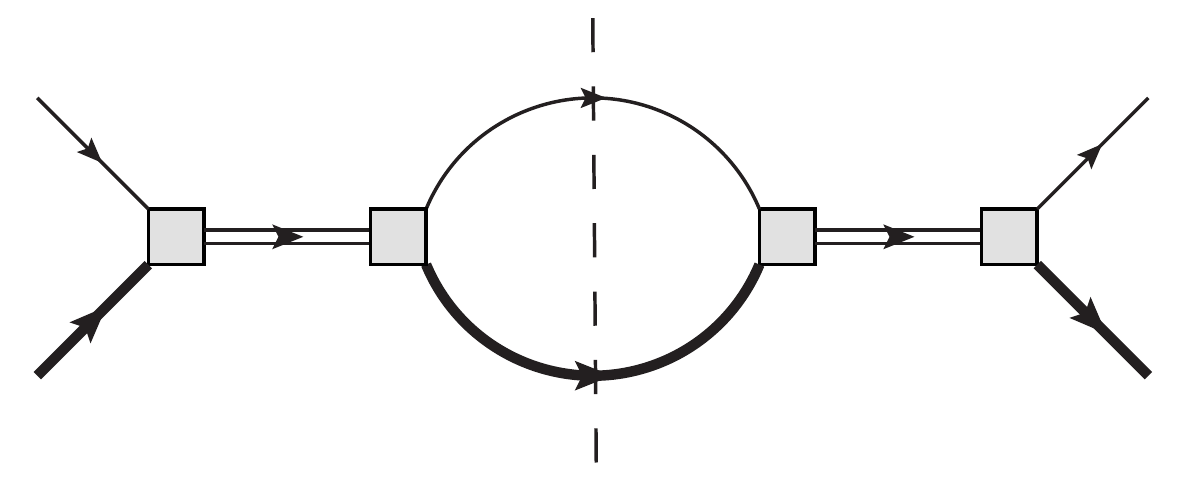}
  \hspace*{2em}
  \includegraphics[height=2cm]{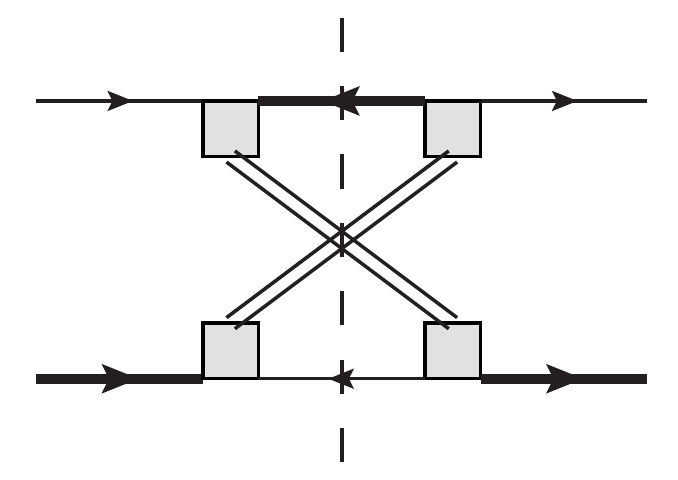}
  \caption{The running of $\bm\lambda$ represented with the auxiliary field propagators and vertices. The double line represents the propagation of any auxiliary field. The dashed line cuts the fermion propagators, whose derivative with respect to $k$ is taken.}
  \label{fig:general_FRG}
\end{figure}

We can repeat the results of the previous subsection with the complete evolution equation \eqref{eq:lambdabarFRG}. With the representation \eqref{eq:lambdarep} we find, just like before:
\begin{equation}
  \delta_{nm} \d_k G^{(H)}_n + (G^{(H)}_n-G^{(H)}_m) \bm v_m^\dagger \d_k\bm v_n
  =  -\bm v_m^\dagger (\d_k\bm{\lambda})\, \bm v_n
\end{equation}
which yields
\begin{equation}
  \d_k G^{(H)}_n = -\bm v_n^\dagger (\d_k\bm{\lambda})\, \bm v_n,\qquad 
  \d_k \bm v_n = \sum_{\ell\neq n} \frac1{G^{(H)}_\ell-G^{(H)}_n} \bm v_\ell(\bm v_\ell^\dagger (\d_k\bm{\lambda})\, \bm v_n).
\end{equation}
Here $\d_k\bm\lambda$ is given by \eqref{eq:lambdabarFRG} where we have to use the representation \eqref{eq:lambdarep}. Diagrammatically we have the contibutions for $\d_k\bm\lambda$ as shown in Fig.~\ref{fig:general_FRG}.
Its matrix elements in the $\bf v_n^\ell$ basis read as
\begin{equation}
   \bm v_m^{\ell\dagger} (\d_k\bm{\lambda})\, \bm v_n^\ell = -G^{(H)}_nG^{(H)}_m \bm
  v_m^{\ell\dagger} (\d_k\bm\G)\, \bm v_n^\ell - L_{mn}^\ell,
\end{equation}
where
\begin{equation}
  L_{mn}^\ell = v^{m\ell*}_{p\alpha\gamma}
  G^{(H)}_{m'} v^{m',\ell+r-q}_{p\alpha\gamma'} v^{m',\ell+r-q*}_{r\beta'\sigma} 
  (\d_k\G)^{\ell+2r-p-q}_{r,\beta'\sigma',\alpha'\gamma'} 
  G^{(H)}_{n'} v^{n',\ell+r-p}_{r\alpha'\gamma} v^{n',\ell+r-p*}_{q\beta\sigma'} 
  v^{n\ell}_{q\beta\sigma}.
\end{equation}
This is a very complicated object containing integration over $p,\, q$ and $r$, therefore it is practically unapproachable if we keep all auxiliary fields.  
However, if we keep only a few of them, this term also simplifies significantly. Thus it may be included in the full analysis of the truncation effects in $n$.

\subsection{Ground state approximation}

One might ask what is the consequence when one truncates (38) to a single spectral component. This truncation is substantially different from introducing a single bosonic field at the level of the effective action, which has been shown in section III not to lead to the correct FRG-equation. Here the correct infinite set of FRG-BS equations is truncated to a single eigenvector.


Keeping the lowest level means that we try to represent the four-point function with a single particle exchanged in the s-channel:
\begin{equation}
  \bm \lambda = c_0\bm P_0,\qquad \bm P_0= \bm v_0\otimes \bm v_0^\dagger.
\end{equation}
Having a look on \eqref{eq:BSFRG} we see that $\bm v_0$ is constant, while for $\K_0=-1/c_0$ we obtain
\begin{equation}
  \d_k \K_0 = - \bm v_0^\dagger\d_k\bm\G \bm v_0
\end{equation}
that has the solution, assuming $\bm\G_\Lambda=0$
\begin{equation}
  \K_{0k} = \K_{0\Lambda} -  \bm v_0^\dagger \bm\G_k \bm v_0.
\end{equation}
The initial condition $\bm\lambda_\Lambda= \bm V$ is hardly can be satisfied, since $\bm\lambda$ is now a pure projector. The best approximation is that we start from $\bm \lambda_\Lambda = \bm P_0 \bm V\bm P_0$, which means $c_0=\bm v_0^\dagger \bm V \bm v_0=-1/\K_\Lambda$. Thus we have 
\begin{equation}
  \K_{0k} = -\frac{ 1+ {\bm v_0^\dagger \bm V \bm v_0} \bm
    v_0^\dagger \bm\G_k \bm v_0}{\bm v_0^\dagger \bm V \bm v_0}.
\end{equation}
The condition to find a bound state thus reads
\begin{equation}
  {\bm v_0^\dagger \bm V \bm v_0} \bm v_0^\dagger \bm\G \bm v_0 =-1
\end{equation}
(where $\bm \G=\bm \G_{k=0}$). This equation should determine the bound state energy. We note, however, that the exact equation would be (c.f. \eqref{eq:BSE0}) $\bm V\bm\G \bm v_0 = -\bm v_0$, or $\bm v_0 \bm V\bm\G \bm v_0=-1$. It is numerically conceivable that the extra projector between $\bm V$ and $\bm \G$ does not count, but in general it is not true. For example, one can construct examples with $2\times2$ matrices where $\bm V$ and $\bm\G$ are hermitean, $\bm V\bm\G \bm v_-=-\bm v_-$, but $\bm v_-^\dagger \bm V\bm v_-=0$ while $\bm v_-^\dagger \bm\G\bm v_-=1$.

A systematic investigation of the truncation of \eqref{eq:BSFRG} is presented in the next section for the exactly solvable problem of non-relativistic electric interaction of charged particles.

\section{FRG solution of the bound state problem of non-relativistic charged fields}

Let us modify the integral in \eqref{eq:int1} in the spirit of FRG, by introducing a regularized momentum as
\begin{equation}
  \label{eq:int2}
  \pint3\p \frac{4\pi}{(\bm q-\bm p)^2(\bar E + p^2_k)} v_k(\bm p) 
  = v_k(\bm q),\qquad p^2_k = \Theta(p-k) p^2 + \Theta(k-p) k^2.
\end{equation}
Note, however, that no regulator is applied in the Fourier transform of the potential.
We see that we recover \eqref{eq:int1} for $k=0$. On the other hand in the formal limit of $k\to\infty$ we obtain an easily solvable system. There, namely, the $p$-integral appears as a convolution, so in real space the above equation reads
\begin{equation}
  k\to\infty:\qquad V(\bm x) v_\infty(\bm x) = (\bar E+k^2) v_\infty(\bm x),
\end{equation}
which has a solution $v_\infty(\bm x)\sim \delta(\bm x-\bm r_0)$. Thus, if we can write down a differential equation for the scale dependence of $v_k({\bf q})$ with $k$-derivatives, we can start from a known exact solution at $k\to\infty$, and arrive at the solution of the actual problem at $k=0$. 

For the object obeying a differential equation in $k$, we introduce the hermitean matrix
\begin{equation}
\label{eq:herm-matr}
  \bm\lambda^{(k,\bar E)} = (1+\bm V\bm\G^{(k,\bar E)})^{-1} \bm V,
\end{equation}
where $\bm\G^{(k,\bar E)}$ is the same as $\bm\G^{(\bar E)}$ with $p\to p_k$ substitution. This choice is dictated by the complete formal analogy of \eqref{eq:herm-matr} with the 4-point function of the relativistic theory (see \eqref{eq:M}). We see from \eqref{eq:herm-matr} that $\bm \lambda^{(k,\bar E)}$ is singular when $\bar E$ is the energy of one of the bound states of the regularized system. For the $k$-evolution of $\bm \lambda^{(k,\bar E)}$ it is easy to set up a differential equation, which is the analogue of the FRG equation. After differentiating \eqref{eq:herm-matr} with respect to $k$ we obtain (using the $k$-independence of $V$)
\begin{equation}
  \label{eq:dkG}	
  \d_k\bm\lambda^{-1} = \d_k\bm\G^{(k,\bar E)} =
  \frac{2k}{(\bar E + k^2)^2}\Theta(k-p)\delta_{pq},
\end{equation}
where the last expression is valid for the Coulomb problem.

For the practical solution we represent the $\bm\lambda$ matrix with its eigenvectors, that is using its spectral decomposition. Taking into account that it is hermitean, we may write
\begin{equation}
  \bm\lambda^{-1}  = \sum_n \K_n^{(k)}\bm x_n^{(k)}\otimes\bm
  x_n^{(k)\dagger}, 
\end{equation}
where $\bm x_n^{(k)}$ are the (orthonormal) eigenvectors and
$\K_n^{(k)}$ are the eigenvalues of $\bm\lambda^{-1}$ at scale $k$. 
Then we have from \eqref{eq:dkG}
\begin{equation}
  \d_k\sum_n \K_n^{(k)}\bm x_n^{(k)}\otimes\bm x_n^{(k)\dagger} = \d_k\bm\G_k.
\end{equation}
To simplify the notation, we will omit the  superscript $(k)$ in the
sequel.

By performing the differentiation, and multiplying the expression by $\bm x_m$ we arrive for the index pairs $m=n$ and $m\neq n$ the respective equations
\begin{eqnarray}
&& \d_k\K_n= \bm x_n^\dagger \d_k\bm\G_k\bm x_n,\nn
&& \d_k\bm x_n = \sum_m \frac{\bm x_m^\dagger \d_k\bm\G_k \bm
    x_n}{\K_n-\K_m} \bm x_m.
\label{eq:rgKx}
\end{eqnarray}
A most remarkable feature of \eqref{eq:rgKx} is its completely identical structure to the general FRG equation \eqref{eq:BSFRG}. This suggests similar convergence features for both cases against the truncation of the spectral basis. 
 The evolution equation of the matrix $\lambda^{-1}$ is replaced now by the evolution of the kernel and the eigenfunctions of its spectral decomposition. The latter correspond to the  the composite fields of the relativistic field theory. The technical steps for solving \eqref{eq:rgKx} are outlined in Appendix \ref{sec:App-1}.

Let us analyze the characteristic features of the solution.

\begin{figure}[htbp]
  \centering
  \includegraphics[height=4.5cm]{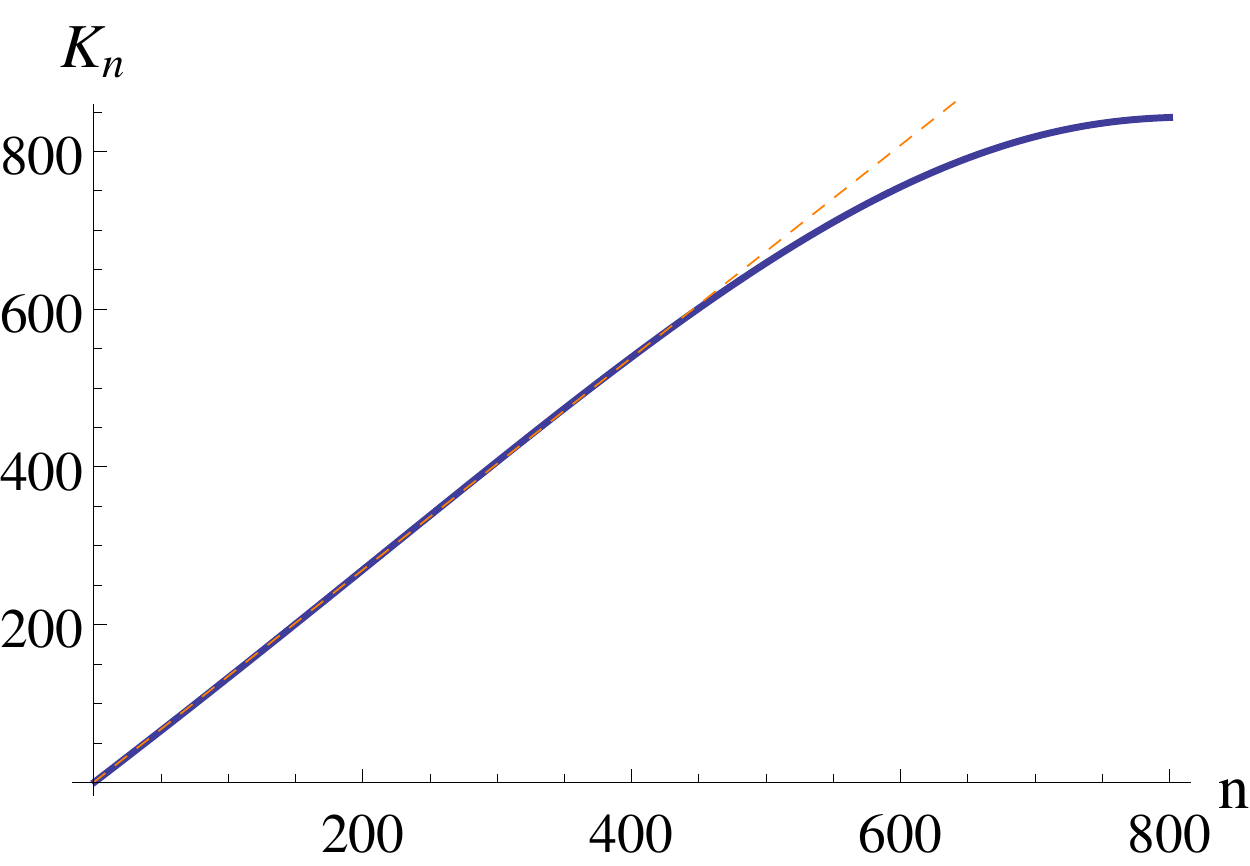}  \qquad
  \includegraphics[height=4.5cm]{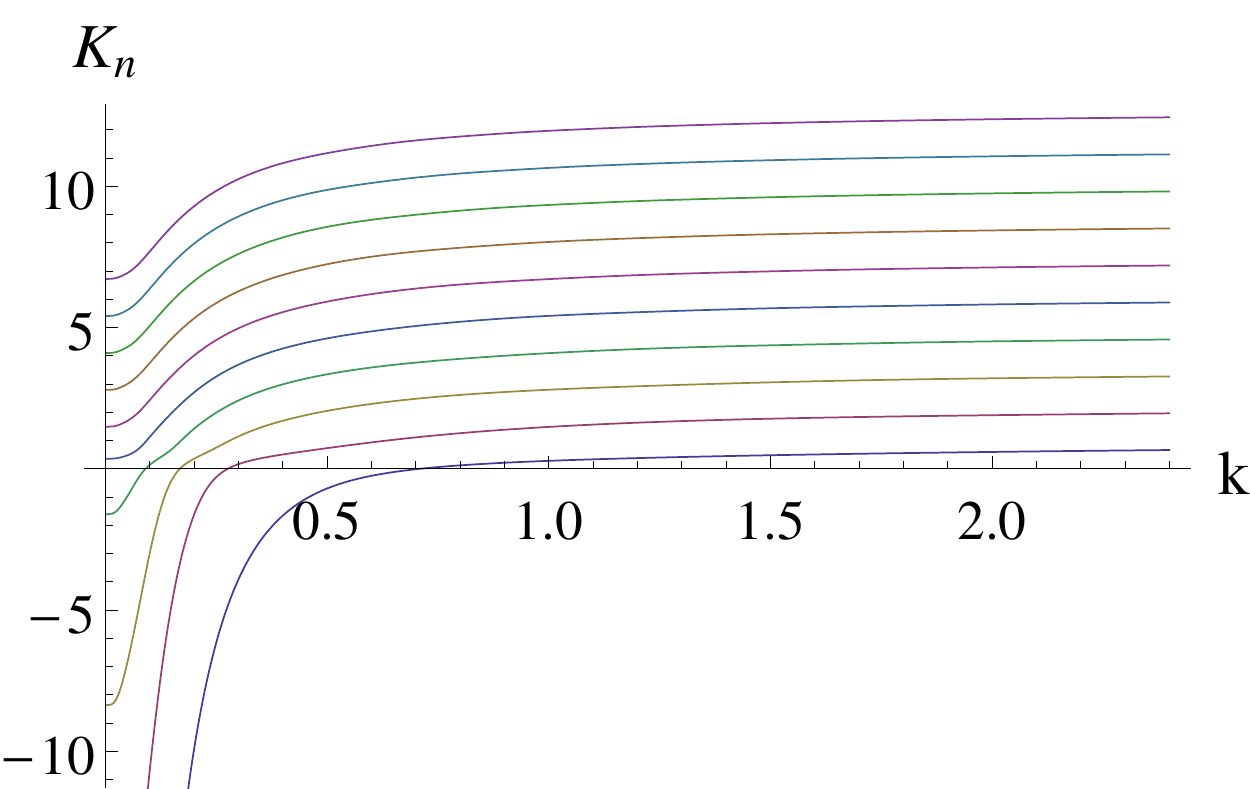}
  \caption{(left) Initial kernel values. (right) The $k$-evolution of    the first 10 energy levels at $\bar E=\frac1{80}$. Since $80>4*16$, the first 4 energy levels cross zero before $k=0$.}
  \label{fig:Kninit}
\end{figure}

The starting $\K_n^{\Lambda}$ eigenvalues \eqref{eq:kinit} are almost linear functions of $n$. In our numerical example $N=800$ and $dp=0.003$ and the initial spectrum can be seen in the left panel of Fig.~\ref{fig:Kninit}.

Starting the RG evolution from these values we can observe the curves shown in the right panel of Fig.~\ref{fig:Kninit}. We see that the values are getting smaller as we decrease the scale, and eventually they cross zero. The bound state energies are coming from the condition that $\K_n(\bar E_n)=0$ at $k=0$.

\begin{figure}[htbp]
  \centering
  \includegraphics[height=4.5cm]{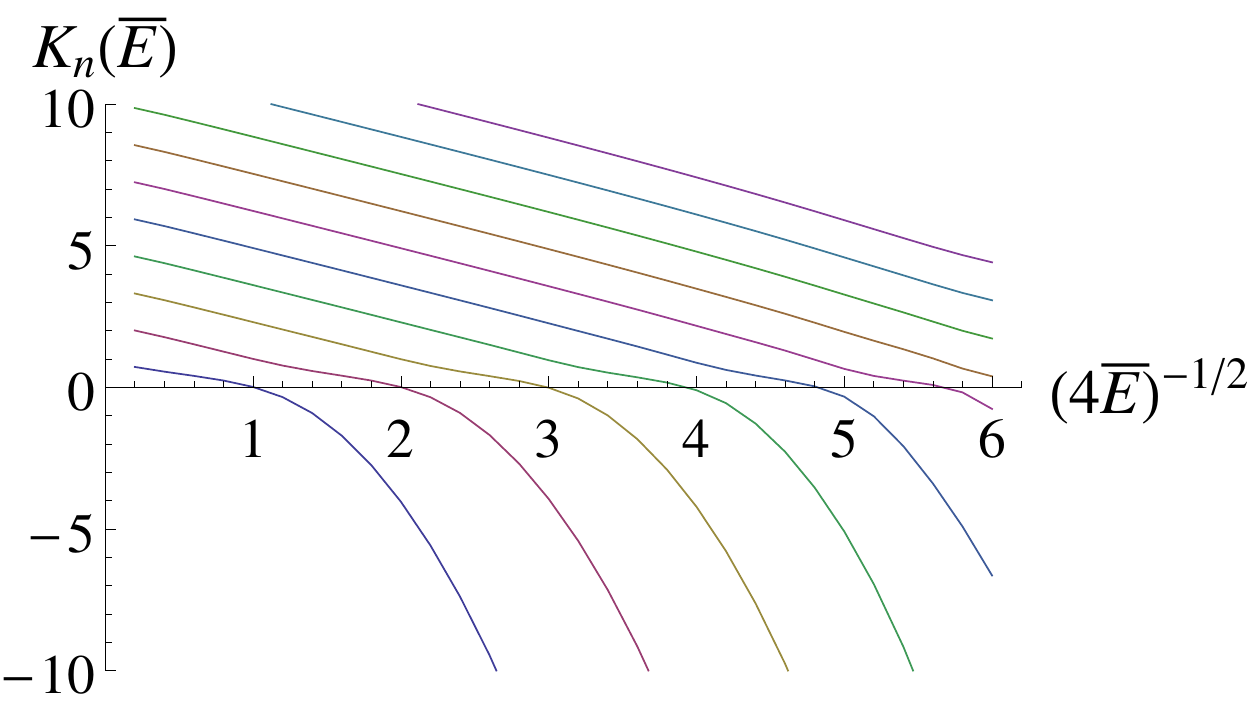}
  \includegraphics[height=4.5cm]{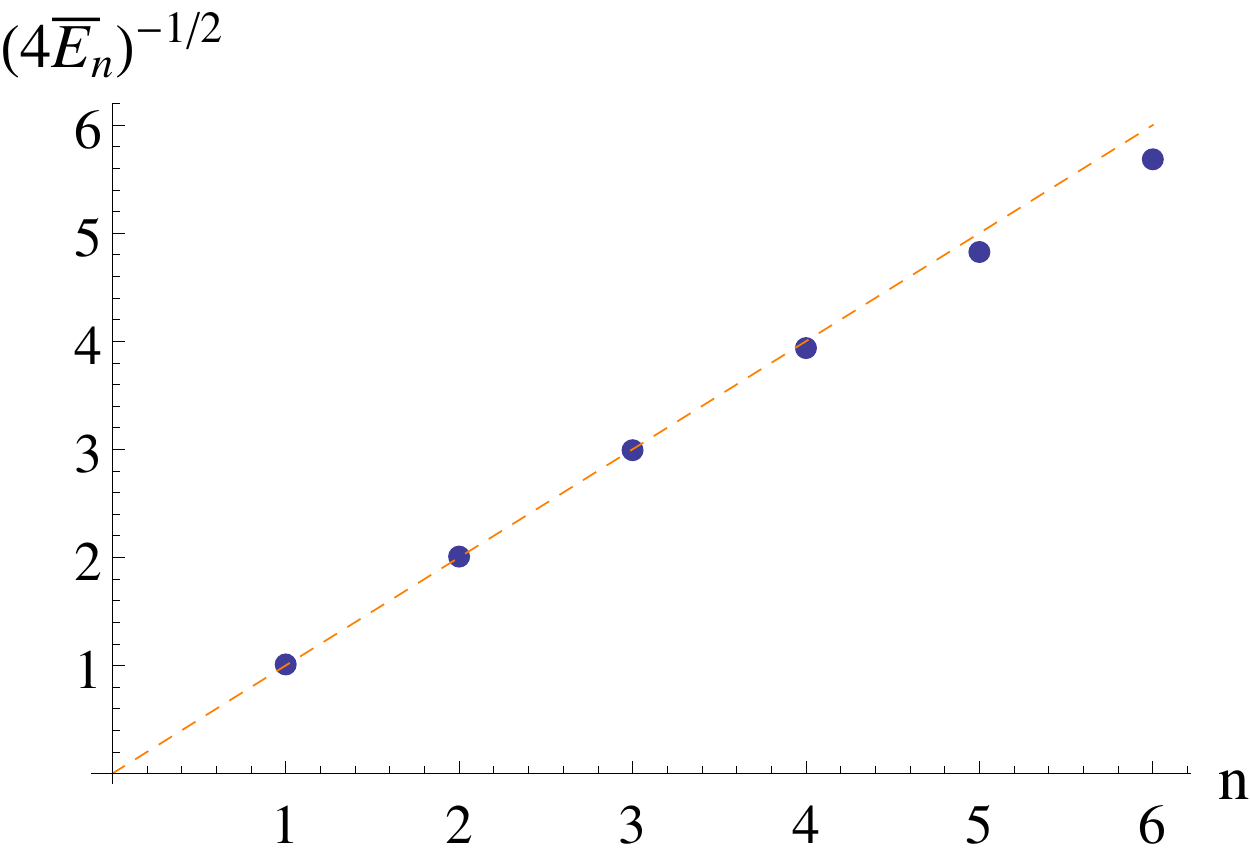}
  \caption{(left) The $N_{eff}$ dependence of the kernel eigenvalues,    where $\bar E=1/(4N_{eff}^2)$. (right) The solutions of $\K_n(\bar E_n)=0$ expressed through $N_{eff}$. The dashed orange line shows the exact result.}
  \label{fig:Escan}
\end{figure}

We can also perform an energy scan: in the left panel of Fig.~\ref{fig:Escan} we display $\K_n(\bar E=\frac1{4N_{eff}^2})$ as a function of $N_{eff}$. We see that the kernel is zero at about $N_{eff}\in$ integers, which is the solution of the Coulomb problem. In the right panel we give the resulting section points.

\begin{figure}[htbp]
  \centering
  \includegraphics[height=4.5cm]{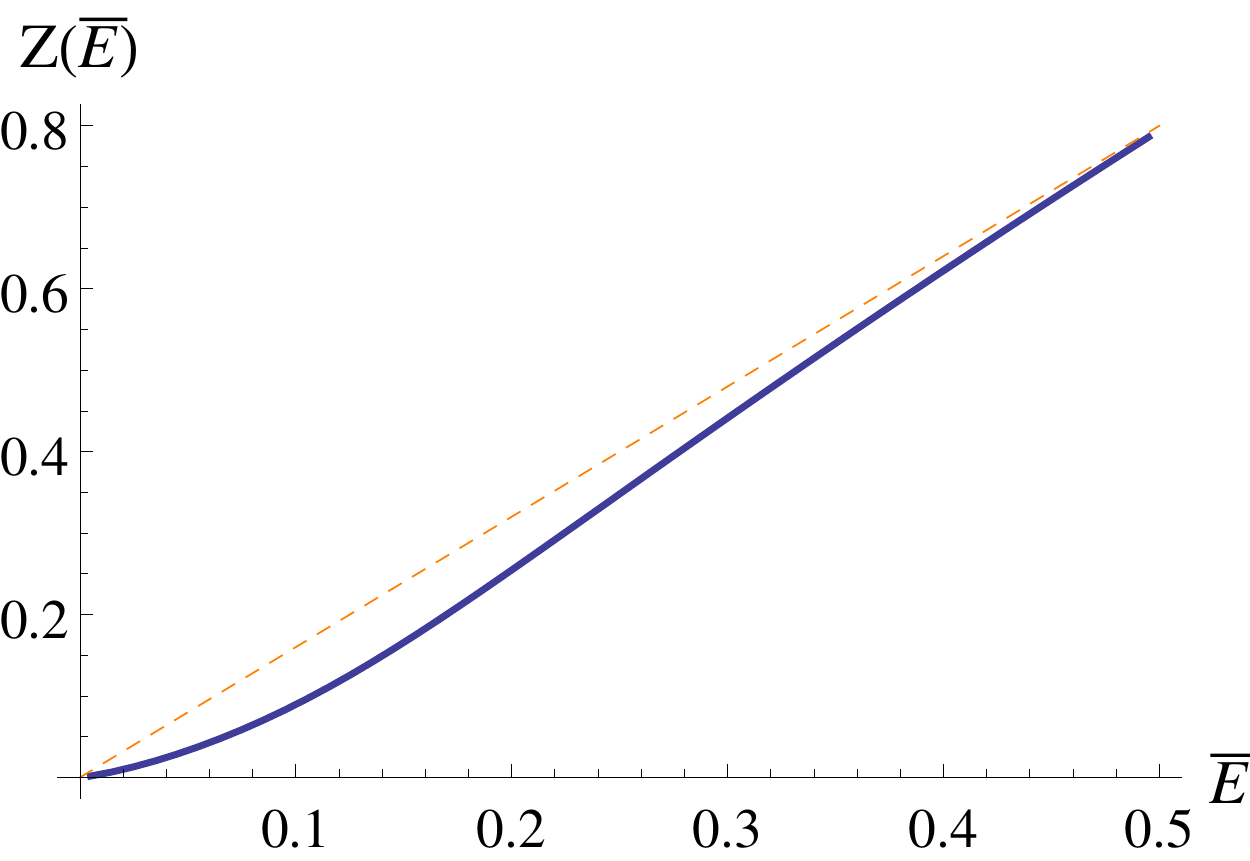}\qquad
  \includegraphics[height=4.5cm]{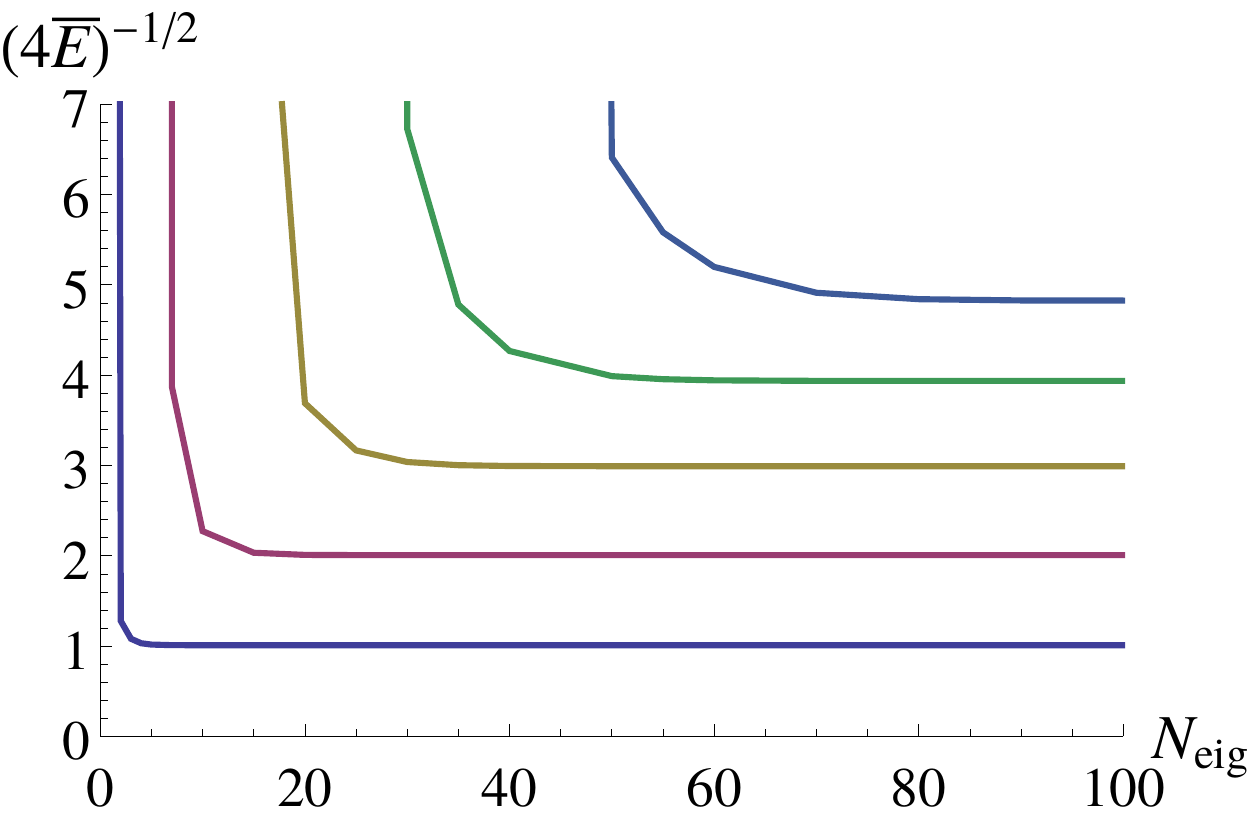}
  \caption{(left panel) The residuum (wave function renormalization) of the propagator of the ground state. (right panel) The estimated energy levels from keeping $N_{eig}$ eigenvectors.}
  \label{fig:ResG}
\end{figure}

The functions $\K_n(\bar E)$ contain, however, more information than just the bound state energies. This function is the bound state kernel, its inverse is the bound state propagator, c.f. Section \ref{sec:introbound}. Considering the ground state, the pole of the propagator is at $\bar E_0\approx 0.245$. If we represent the propagator as
\begin{equation}
  \label{eq:G0}
  G_0(\bar E) = \frac{Z(\bar E)}{\bar E-\bar E_0},
\end{equation}
then the residuum (wave function renormalization, { equivalent in non-relativistic quantum mechanics to the absolute square of the ground state wave function}) is shown in the left panel of Fig.~\ref{fig:ResG}.

These are all interesting features, but the most important message concerns the necessary number of eigenvectors for a reliable estimate of the bound state energies. In the right panel of Fig.~\ref{fig:ResG} we show the estimated energy levels by keeping $N_{eig}$ eigenvectors. It is remarkable that the ground state can be estimated with 8\% precision using only three eigenvectors, and for a precision of 28\% it is enough to consider just the first two eigenvectors. This means that we can effectively reduce the dimension of the Hilbert space, concentrating only on the first few eigenvectors (cf. also TCSH approximations \cite{Yurov:1989yu, Bajnok:2015bgw}). It is also important, that keeping only the ground state is not enough, it leads to a complete misidentification of the ground state energy.

\section{Conclusions}
\label{sec:concl}

The problem of bound states constituted by two different fermions can be solved by searching for the poles of the 4-fermion function in the appropriate channel. In principle, any theory where the interaction between the fermi-fields is mediated by the exchange of some force-field and/or the constituents themselves can be mapped on a pure fermi theory with nonlocal 4-fermi interaction, and the bound-state problem can be solved in this framework.

A very attractive approach to this problem consists of introducing composite fields into the effective action representing the bound states and reducing the investigation of the 4-fermi function to that of 2-composite propagator function.

In this paper first we have formulated the equivalence criterium of the two approaches in quantitative terms (section II). The conclusion was that the conditions can be fullfilled by introducing infinite number of composite fields with appropriate nonlocal Yukawa couplings to the constituents. The functional correspondence established between the two formulation has allowed us to write down the exact equations of the functional renormalisation group evolution for the composite fields exploiting the FRG equations of the 4-fermi theory (section IV). The Bethe-Salpeter method for finding relativistic bound states has been shown to follow from the so-called s-channel approximation to the FRG-equations.
  
On the algorithmic side it was pointed out that the reduced set of FRG-equations resulting from truncating the number of composite fields at some finite number makes sense, and the accuracy of this approximation can be tested by varying the dimensionality of the system of FRG equations in a systematic way. This type of study has been illustrated in section V through the example of the non-relativistic bound state of two oppositely charged fermi fields (“H-atom” problem).
However, the restriction to a single composite field on the level of the definition of the effective action was shown to lead to an FRG equation incompatible with the Bethe-Salpeter equation, namely these FRG equations do not produce any pole in the 3-point function connecting the composite field to the constituting two fermi fields (section III).

The next step of developing a consistent field theory of the composite fields is the application of the proposed framework to relativistic fermion bound states in specific models. The interesting question if one finds bound states in the symmetric phase of the chiral Nambu—Jona- Lasinio model has been studied recently by us based on the solution of the FRG-equations derived for a large subset of bosonic composite fields and extracting a potential between the constituents by comparing the infrared limiting values obtained for the energy of interaction for components of different total squared mass. \cite{Jakovac:2019zzw}

\bibliography{BS-rev}

\begin{thebibliography}{10}

\bibitem{Salpeter:1951sz}
E.~E. Salpeter and H.~A. Bethe, ``{A Relativistic equation for bound state
  problems},'' {\em Phys. Rev.}, vol.~84, pp.~1232--1242, 1951.

\bibitem{Itzykson:1980rh}
C.~Itzykson and J.~B. Zuber, {\em {Quantum Field Theory}}.
\newblock International Series In Pure and Applied Physics, New York:
  McGraw-Hill, 1980.

\bibitem{Maris:1997piK}
P.~Maris and C.~Roberts, ``{pi- and K-meson Bethe-Salpeter amplitudes},'' {\em
  Phys. Rev.}, vol.~C56, no.~3, pp.~3369--3383, 1997.

\bibitem{Maris:1999vmd}
P.~Maris and P.C.Tandy, ``{Bethe Salpeter study of vector meson masses and
  decay contants},'' {\em Phys. Rev.}, vol.~C60, no.~3, p.~055214, 1999.

\bibitem{Hilger:2014nma}
T.~Hilger, C.~Popovici, M.~Gomez-Rocha, and A.~Krassnigg, ``{Spectra of heavy
  quarkonia in a Bethe-Salpeter-equation approach},'' {\em Phys. Rev.},
  vol.~D91, no.~3, p.~034013, 2015.

\bibitem{Alkofer:2000wg}
R.~Alkofer and L.~von Smekal, ``{The Infrared behavior of QCD Green's
  functions: Confinement dynamical symmetry breaking, and hadrons as
  relativistic bound states},'' {\em Phys. Rept.}, vol.~353, p.~281, 2001.

\bibitem{Sanchis-Alepuz:2015tha}
H.~Sanchis-Alepuz and R.~Williams, ``{Hadronic Observables from Dyson-Schwinger
  and Bethe-Salpeter equations},'' {\em J. Phys. Conf. Ser.}, vol.~631, no.~1,
  p.~012064, 2015.

\bibitem{Eichmann:2016yit}
G.~Eichmann, H.~Sanchis-Alepuz, R.~Williams, R.~Alkofer, and C.~S. Fischer,
  ``{Baryons as relativistic three-quark bound states},'' {\em Prog. Part.
  Nucl. Phys.}, vol.~91, pp.~1--100, 2016.

\bibitem{Hilger:2017jti}
T.~Hilger, M.~Gómez-Rocha, A.~Krassnigg, and W.~Lucha, ``{Aspects of
  open-flavour mesons in a comprehensive DSBSE study},'' {\em Eur. Phys. J.},
  vol.~A53, no.~10, p.~213, 2017.

\bibitem{Watson:2004kd}
P.~Watson, W.~Cassing, and P.~C. Tandy, ``{Bethe-Salpeter meson masses beyond
  ladder approximation},'' {\em Few Body Syst.}, vol.~35, pp.~129--153, 2004.

\bibitem{sanchis-alepuz:2015plb}
H.~Sanchis-Alepuz and R.~Williams, ``{From Quarks and Gluons to Hadrons: Chiral
  Symmetry Breaking in Dynamical QCD},'' {\em Phys. Lett.}, vol.~749,
  pp.~592--596, 2015.

\bibitem{williams:2016prd}
R.~Williams, C.~S. Fischer, and W.~Heupel, ``{Light mesons in QCD and
  unquenching effects from the 3PI effective action},'' {\em Phys. Rev.},
  vol.~D93, p.~034026, 2016.

\bibitem{Wetterich:1992yh}
C.~Wetterich, ``{Exact evolution equation for the effective potential},'' {\em
  Phys. Lett.}, vol.~B301, pp.~90--94, 1993.

\bibitem{Morris:1993qb}
T.~R. Morris, ``{The Exact renormalization group and approximate solutions},''
  {\em Int. J. Mod. Phys.}, vol.~A9, pp.~2411--2450, 1994.

\bibitem{gies:2012lnp}
H.~Gies, ``{Introduction to the functional RG and applications to gauge
  theories},'' {\em Lect. Notes Phys.}, vol.~852, p.~034026, 2012.

\bibitem{braun:2012jpg}
J.~Braun, ``{Fermion Interactions and Universal Behavior in Strongly
  Interacting Theories },'' {\em J.Phys.}, vol.~G39, p.~033001, 2012.

\bibitem{Ellwanger:1993mw}
U.~Ellwanger, ``{FLow equations for N point functions and bound states},'' {\em
  Z. Phys.}, vol.~C62, pp.~503--510, 1994.
\newblock [,206(1993)].

\bibitem{Ellwanger:1994wy}
U.~Ellwanger and C.~Wetterich, ``{Evolution equations for the quark - meson
  transition},'' {\em Nucl. Phys.}, vol.~B423, pp.~137--170, 1994.

\bibitem{Blaizot:2005xy}
J.~P. Blaizot, R.~Mendez~Galain, and N.~Wschebor, ``{A New method to solve the
  non perturbative renormalization group equations},'' {\em Phys. Lett.},
  vol.~B632, pp.~571--578, 2006.

\bibitem{Benitez:2011xx}
F.~Benitez, J.~P. Blaizot, H.~Chate, B.~Delamotte, R.~Mendez-Galain, and
  N.~Wschebor, ``{Non-perturbative renormalization group preserving
  full-momentum dependence: implementation and quantitative evaluation},'' {\em
  Phys. Rev.}, vol.~E85, p.~026707, 2012.

\bibitem{Blaizot:2005wd}
J.-P. Blaizot, R.~Mendez-Galain, and N.~Wschebor, ``{Non perturbative
  renormalisation group and momentum dependence of n-point functions (I)},''
  {\em Phys. Rev.}, vol.~E74, p.~051116, 2006.

\bibitem{Blaizot:2006vr}
J.-P. Blaizot, R.~Mendez-Galain, and N.~Wschebor, ``{Non perturbative
  renormalization group and momentum dependence of n-point functions. II.},''
  {\em Phys. Rev.}, vol.~E74, p.~051117, 2006.

\bibitem{rose16}
F.~Rose, F.~Benitez, F.~L\'eonard, and B.~Delamotte, ``{Bound states of the
  $\Phi^4$ model via the nonperturbative renormalization group},'' {\em Phys.
  Rev.}, vol.~D93, no.~12, p.~125018, 2016.

\bibitem{Caselle:2000yx}
M.~Caselle, M.~Hasenbusch, P.~Provero, and K.~Zarembo, ``{Bound states in the
  three-dimensional phi**4 model},'' {\em Phys. Rev.}, vol.~D62, p.~017901,
  2000.

\bibitem{Caselle:2001im}
M.~Caselle, M.~Hasenbusch, P.~Provero, and K.~Zarembo, ``{Bound states and
  glueballs in three-dimensional Ising systems},'' {\em Nucl. Phys.},
  vol.~B623, pp.~474--492, 2002.

\bibitem{Jungnickel:1995fp}
D.~U. Jungnickel and C.~Wetterich, ``{Effective action for the chiral
  quark-meson model},'' {\em Phys. Rev.}, vol.~D53, pp.~5142--5175, 1996.

\bibitem{Gies:2001nw}
H.~Gies and C.~Wetterich, ``{Renormalization flow of bound states},'' {\em
  Phys. Rev.}, vol.~D65, p.~065001, 2002.

\bibitem{Pawlowski:2005xe}
J.~M. Pawlowski, ``{Aspects of the functional renormalisation group},'' {\em
  Annals Phys.}, vol.~322, pp.~2831--2915, 2007.

\bibitem{Floerchinger:2009uf}
S.~Floerchinger and C.~Wetterich, ``{Exact flow equation for composite
  operators},'' {\em Phys. Lett.}, vol.~B680, pp.~371--376, 2009.

\bibitem{Floerchinger:2010da}
S.~Floerchinger, ``{Exact Flow Equation for Bound States},'' {\em Eur. Phys.
  J.}, vol.~C69, pp.~119--132, 2010.

\bibitem{Alkofer:2018guy}
R.~Alkofer, A.~Maas, W.~A. Mian, M.~Mitter, J.~Paris-Lopez, J.~M. Pawlowski,
  and N.~Wink, ``{Bound state properties from the Functional Renormalisation
  Group},'' 2018.

\bibitem{Kamikado2014}
K.~Kamikado, N.~Strodthoff, L.~von Smekal, and J.~Wambach, ``{Real-time
  correlation functions in the $O(N)$ model from the functional renormalization
  group},'' {\em Eur. Phys. J.}, vol.~C74, no.~3, p.~2806, 2014.

\bibitem{tripolt14}
R.-A. Tripolt, L.~von Smekal, and J.~Wambach, ``{Flow equations for spectral
  functions at finite external momenta},'' {\em Phys. Rev.}, vol.~D90, no.~7,
  p.~074031, 2014.

\bibitem{wambach14}
J.~Wambach, R.-A. Tripolt, N.~Strodthoff, and L.~von Smekal, ``{Spectral
  Functions from the Functional Renormalization Group},'' {\em Nucl. Phys.},
  vol.~A928, pp.~156--167, 2014.

\bibitem{Yurov:1989yu}
V.~P. Yurov and A.~B. Zamolodchikov, ``{TRUNCATED CONFORMAL SPACE APPROACH TO
  SCALING LEE-YANG MODEL},'' {\em Int. J. Mod. Phys.}, vol.~A5, pp.~3221--3246,
  1990.

\bibitem{Bajnok:2015bgw}
Z.~Bajnok and M.~Lajer, ``{Truncated Hilbert space approach to the 2d
  $\phi^{4}$ theory},'' {\em JHEP}, vol.~10, p.~050, 2016.

\bibitem{Jaeckel:2002rm}
J.~Jaeckel and C.~Wetterich, ``{Flow equations without mean field ambiguity},''
  {\em Phys. Rev.}, vol.~D68, p.~025020, 2003.

\bibitem{braun:2016dh}
J.~Braun, L.~Fister, J.~M. Pawlowski, and F.~Rennecke, ``{From Quarks and
  Gluons to Hadrons: Chiral Symmetry Breaking in Dynamical QCD},'' {\em Phys.
  Rev.}, vol.~D94, p.~034016, 2016.

\bibitem{Jakovac:2019zzw}
A.~Jakovac and A.~Patkos, ``{Interacting two-particle states in the symmetric
  phase of the chiral Nambu--Jona-Lasinio model},'' 2019.

\end{thebibliography}
\bibliographystyle{ieeetr}

\appendix

\section{Bethe-Salpeter resummation and its application to non-relativistic QED}
\label{sec:BSE}

The central quantity in the study of the bound states is the
$\exv{\psi\psi^\dagger\chi\chi^\dagger}$ 4-point function. Since we
expect that the bound state consists of a $\psi$ and a $\chi$ particle,
we can find it as a pole in the $\psi$-$\chi$ scattering amplitude.

\subsection{Four point function at one loop level}

First we will compute the four point function in perturbation theory
at one-loop level. The four point function depends on four momenta,
but the energy-momentum conservation reduces the number of independent
variables to three. For later convenience we will define the amputed
4-point amplitude as
\begin{equation}
  \label{eq:4fermamp}
  \M_{p\alpha\gamma,q\beta\sigma}^\ell := -\exv{ \hat\psi^{\dagger}_{\alpha p}
    \hat\psi_{\beta q}\hat\chi^\dagger_{\gamma,\ell-p} \hat\chi_{\sigma, \ell-q}
  }\biggr|_{amputed} 
\end{equation}
where the hat means that that line is amputed, and the indices refer
to the amputed ends. The quantity $\ell$ denotes the total incoming
momentum. The minus sign is here for convenience, it signals that
the interaction is attractive. 

The indices on the left hand side are arranged in a way that will be convenient to use later on.
We work with matrices with (multi)indices $p\alpha\beta$.
Relative to these multi-indices the above defined object is hermitean:
\begin{equation}
  \M_{p\alpha\gamma,q\beta\sigma}^{\ell*} = \M_{q\beta\sigma,p\alpha\gamma}^{\ell}.
\end{equation}

\begin{figure}[htbp]
  \centering
  \includegraphics[height=1.8cm]{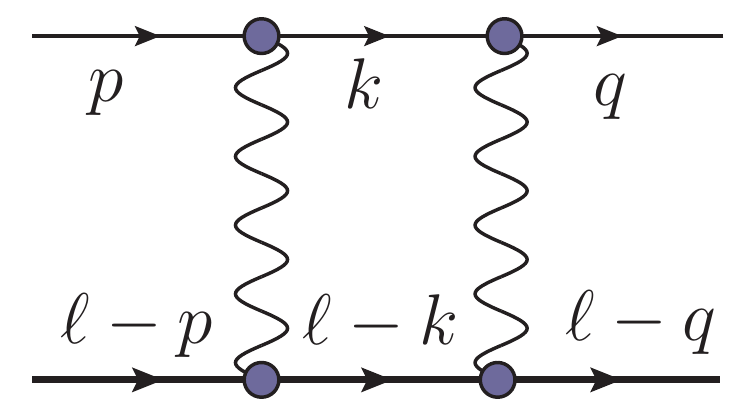}
  \hspace*{2em}
  \includegraphics[height=1.8cm]{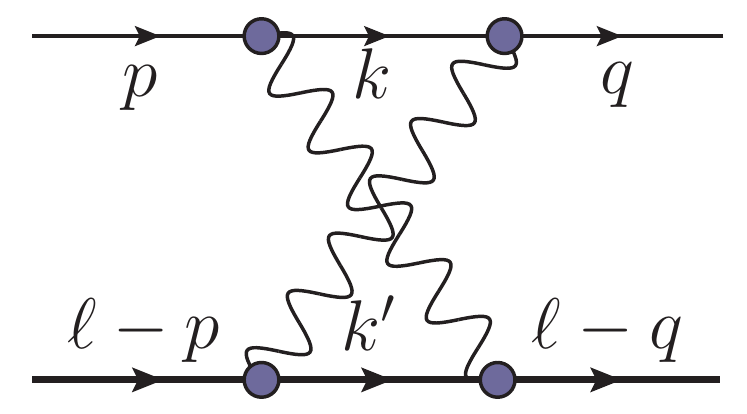}
  \caption{The 1PI one-loop Feynman diagrams contributing to the
    connected 4-point function ($k'=\ell-p+k-q$). Thin lines represent
    $\psi$, thick lines $\chi$, and curly lines stand for $A$
    propagators.}
  \label{fig:fourpf_1PI}
\end{figure}

At tree level we have a single diagram contributing (cf. Fig.~\ref{fig:fourpf_tree}) which in the non-relativistic limit corresponds to the electrostatic potential between the two fermions.
At one loop level we show the two contributing 1PI diagrams (cf. Fig.~\ref{fig:fourpf_1PI}) with the analytic expressions
\begin{eqnarray}
  \label{eq:Moneloop}
  (\M_{p\alpha\gamma,q\beta\sigma}^\ell)_{1PI}^{1-loop} = -\pint4k\biggl[
  && (\gamma^a_{pk}G^{(\psi)}_{k}\gamma^b_{kq})_{\alpha\beta}
     (\gamma^a_{\ell-p,\ell-k} G^{(\chi)}_{\ell-k}
     \gamma^b_{\ell-k,\ell-q})_{\gamma\sigma} G^{(A)}_{p-k} G^{(A)}_{k-q}
      +\nn
  && + (\gamma^a_{pk}G^{(\psi)}_{k}\gamma^b_{kq})_{\alpha\beta}
     (\gamma^b_{\ell-p,\ell-p+k-q} G^{(\chi)}_{\ell-p+k-q}
     \gamma^a_{\ell-p+k-q,\ell-q})_{\gamma\sigma} G^{(A)}_{p-k} G^{(A)}_{k-q}\biggr],
\end{eqnarray}
where we have suppressed the internal spinor indices for
readability.

\begin{figure}[htbp]
  {
  \begin{picture}(280,50)
    \put(0,21){\mbox{$\bar V=$}}
    \put(25,0){\includegraphics[height=1.8cm]{fourpf-tree}}
    \put(90,22){$+$}
    \put(100,0){\includegraphics[height=1.8cm]{fourpf-1loop2}}
    \put(195,22){$\equiv$}
    \put(210,0){\includegraphics[height=1.8cm]{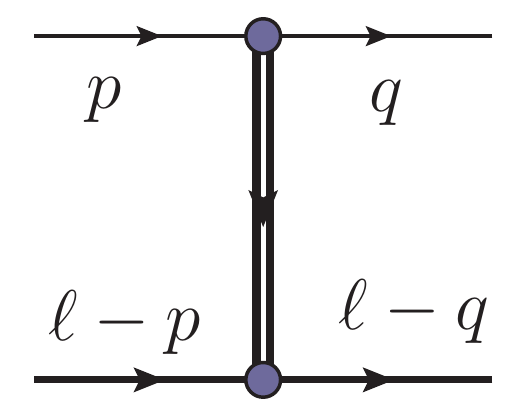}}
  \end{picture}}
  \caption{The diagrams contributing to the effective ``rung'' of the effective ladder shown as a double thick line in the sequel.}
  \label{fig:Vdef}
\end{figure}

In the resummation which leads to BSE (the BS resummation) the crossed leg contribution is included as
a perturbative correction into the potential $V^{(0)}$. One writes for the complemented "potential" denoted by $\bar V$
\begin{eqnarray}
  \bar V_{p\alpha\gamma,q\beta\sigma}^\ell
  =V^{\ell}_{p\alpha\gamma,q\beta\sigma}
  -\pint4k(\gamma^a_{pk}G^{(\psi)}_{k}\gamma^b_{kq})_{\alpha\beta} 
  (\gamma^b_{\ell-p,\ell-p+k-q} G^{(\chi)}_{\ell-p+k-q}
  \gamma^a_{\ell-p+k-q,\ell-q})_{\gamma\sigma} G^{(A)}_{p-k} G^{(A)}_{k-q}.
\end{eqnarray}
The graphic symbol expressing this definition appears in Fig.~\ref{fig:Vdef}.
To be truly consistent, the vertices and the propagators here should
also be dressed. We do not elaborate on this question, which is just a technical side problem from the point of view of the bound state formation.

In addition to the interaction "potential" we make use also of the 2-fermion propagator
\begin{equation}
  \G_{p\alpha\gamma,q\beta\sigma}^\ell = \delta_{pq} G^{(\psi)}_{p,\alpha\beta}
  G^{(\chi)}_{\ell-p,\gamma\sigma}.
\end{equation}
In the multi-index notation this is also hermitean
$\G_{p\alpha\gamma,q\beta\sigma}^{\ell*}=\G_{q\beta\sigma,p\alpha\gamma}^\ell$.
With these notations, at one-loop level in the effective "potential" we have
\begin{equation}
  \M_{p\alpha\gamma,q\beta\sigma}^\ell =
  \bar V_{p\alpha\gamma,q\beta\sigma}^\ell -
  \sum_{\alpha'\gamma'\beta'\sigma'} \pint4k\frac{d^4k'}{(2\pi)^4}
  \bar V_{p\alpha\gamma,k\alpha'\gamma'}^\ell
  \G_{k\alpha'\gamma',k'\beta'\sigma'}^\ell
  \bar V_{k'\beta'\sigma',q\beta\sigma}^\ell. 
\end{equation}
This expression can be written in a formally simpler way using a scalar
product notation. For some quantities $f_{p\alpha\gamma}$ and $g_{p\alpha\gamma}$ the
scalar product is defined as
\begin{equation}
  fg = \sum_{\alpha\gamma} \pint4p f_{p\alpha\gamma}g_{p\alpha\gamma}.
\end{equation}
Then we find
\begin{equation}
  {\bm\M}^\ell = \bm {\bar V}^\ell - {\bm{\bar V}}^\ell{\bm\G}^\ell{\bm{\bar V}}^\ell.
\end{equation}
We see that with these definitions the index $\ell$ is just a spectator index (it
would not be, if we would not include the crossed-leg contribution into the potential). In the following, as far it does not lead to
misunderstanding, we will suppress the explicit reference to $\ell$.

\subsection{The Bethe-Salpeter resummation}

\begin{figure}[htbp]
  {
  \begin{picture}(450,50)
    \put(0,21){\mbox{$\M=$}}
    \put(25,0){\includegraphics[height=1.8cm]{fourpf-Vrep}}
    \put(90,22){$+$}
    \put(100,0){\includegraphics[height=1.8cm]{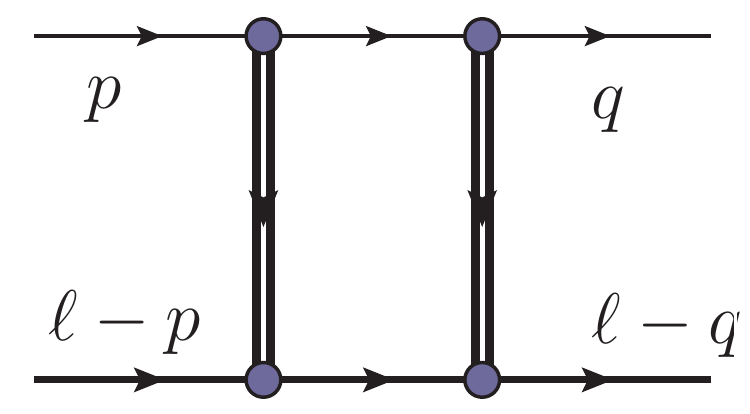}}
    \put(198,22){$+$}
    \put(210,0){\includegraphics[height=1.8cm]{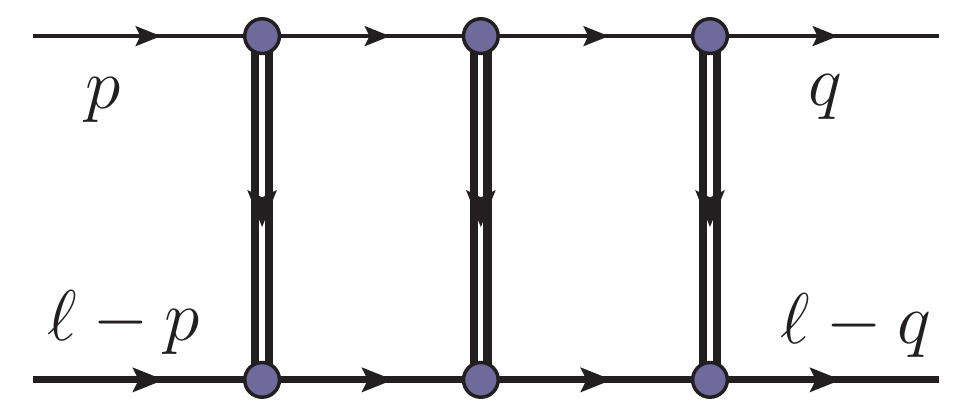}}
    \put(330,22){$+\dots\; \equiv$}
    \put(370,0){\includegraphics[height=1.8cm]{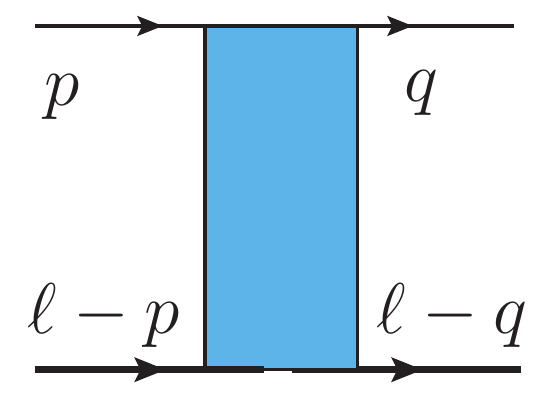}}
  \end{picture}}
  \caption{The BS resummation: only the ladder diagrams are resummed.}
  \label{fig:BSresum}
\end{figure}

At higher order there are several diagrams that contribute. In the
Bethe-Salpeter approximation only the ladder diagrams are taken into
account
\begin{equation}
  \label{eq:M}
  {\bm\M} = {\bm{\bar V}} - {\bm{\bar V}}{\bm\G}{\bm{\bar V}} +
  {\bm{\bar V}}{\bm\G}{\bm{\bar V}}{\bm\G}{\bm{\bar V}} +\dots
  = (1+{\bm{\bar V}} \bm\G)^{-1}{\bm{\bar V}} .
\end{equation}
In graphical representation we see the result in Fig.~\ref{fig:BSresum}.
The 4-point function, as we see, has poles, where the matrix cannot
be inverted. Since the same ladder diagram appears in the resummation of the bound-state -- constituents 3-point function a pole singularity shows up also in that object.

\begin{figure}[htbp]
  {
  \begin{picture}(270,50)
    \put(0,0){\includegraphics[height=1.8cm]{fourpf-BSkernel}}
    \put(80,22){$=$}
    \put(90,0){\includegraphics[height=1.8cm]{fourpf-Vrep}}
    \put(160,22){$+$}
    \put(170,0){\includegraphics[height=1.8cm]{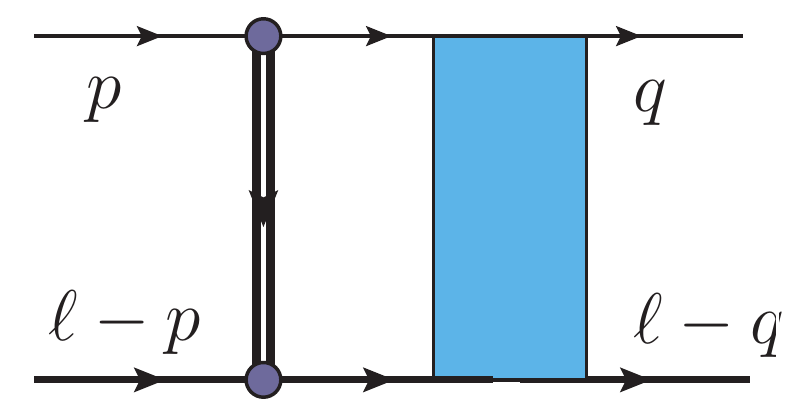}}
  \end{picture}}
  \caption{The BS resummation: only the ladder diagrams are resummed.}
  \label{fig:BSselfcons}
\end{figure}

The result of the resummation \eqref{eq:M} can be also understood as solution of the matrix
equation 
\begin{equation}
  \bm\M = {\bm{\bar V}} - {\bm{\bar V}} \bm\G \bm\M,\qquad
\end{equation}
as it is demonstrated in Fig.~\ref{fig:BSselfcons}.

All these relations contain matrix inversion which does not exists
when the matrix has a zero eigenvalue (then the determinant is also
zero). So the condition of having a pole in the four point function
is equivalent that there exists a vector $\bm u$ that
\begin{equation}
  \label{eq:BSE0}
  {\bm{\bar V}}\bm \G \bm u= -\bm u.
\end{equation}
This is the \emph{Bethe-Salpeter equation}.  Note that in the complete
notation $\bm u$ has several indices: $u^\ell_{p\alpha\gamma}$.

We remark finally that we can write up the resummed 1PI proper 4-fermion
vertex, too:
\begin{equation}
  \bm\Gamma^{(4)} = \bm\M - {\bm{\bar V}} = -(1+{\bm{\bar V}}\bm\G)^{-1}{\bm{\bar V}}\bm\G{\bm{\bar V}}.
\end{equation}

\subsection{Bethe-Salpeter form of the non-relativistic electromagnetic bound state problem }

The Schr\"odinger equation for an electron in the Coulomb
potential can be cast into the form of a Bethe-Salpeter equation in a few steps. The Schr\"odinger equation reads
\begin{equation}
  \label{eq:SCH1}
  {\triangle} \Psi + \frac1x \Psi = \bar E \Psi,
\end{equation}
where we introduced $r_0=\frac{2\hbar^2 \pi\ep_0}{m e^2}$, $\bm x=\frac{\bm r}{r_0}$, and $\bar E =-\frac{2m}{\hbar^2} r_0^2 E$. We concentrate only on the $s$-states, i.e. rotational invariant states ($\ell=0$). Then the above equation simplifies to
\begin{equation}
  \label{eq:Schroed}
  \frac{d^2u}{dx^2} + \frac1{x}u  = \bar E u.
\end{equation}
where $u(x)=x\Psi(x)$. From the analytic solution we know that the energy eigenvalues are
\begin{equation}
  \bar E_n = \frac1{4n^2},\qquad n=1,2,\dots.
\end{equation}
This is the standard approach to the quantum mechanical bound states: we look for the eigensystem of the Hamiltonian, that provides the quantized energy values. 
As a first step towards the BS-form we perform Fourier-transform on \eqref{eq:SCH1}, and introduce $v(\bm q) = (\bar E+q^2) \Psi(\bm q)$. We obtain
\begin{equation}
  \label{eq:int1}
  \pint3\p \frac{4\pi}{(\bm q-\bm p)^2(\bar E + p^2)} v(\bm p) = v(\bm q).
\end{equation}
For the $s$-states, with $\eta(q)=qv(q)$ we find
\begin{equation}
  \eta(q) = \frac1\pi \int\limits_0^\infty\! dp\, \frac{1} {\bar E +
    p^2} \log\left|\frac{q+p}{q-p}\right|\,\eta(p).
\end{equation}
This is an integral equation, which is seemingly more complicated than the original differential equation \eqref{eq:Schroed}. However, introducing the "matrices" with two spatial momenta as indices:
\begin{equation}
  \bm V_{pq} = \frac{4\pi}{(\bm p-\bm q)^2},\qquad
  \bm\G_{pq}^{(\bar E)}=-\frac1{\bar E + p^2}\delta_{pq},
\end{equation}
then \eqref{eq:int1} can be written as
\begin{equation}
  \label{eq:BS}
  \bm V\bm\G^{(\bar E)} \bm v= -\bm v.
\end{equation}
This is formally equivalent to the Bethe-Salpeter equations \eqref{eq:BSE0} (in fact it is the nonrelativistic limit of \eqref{eq:BSE0}).
On the other hand, the eigenvectors of this system clearly correspond to the bound state wave functions.

\section{Solution of the non-relativistic flow equation}
\label{sec:App-1}

Here the technical steps for solving the equations in \eqref{eq:rgKx} are described.
In case of the $s$-state Coulomb problem we may introduce
\begin{equation}
  \label{eq:etaCdefs}
  \eta_n(p)= px_n(p),\qquad\mathrm{and}\qquad
  C_{nm}= \frac1{2\pi^2} \int\limits_0^k\!dq\,\eta_n^*(q)\eta_m(q),
\end{equation}
then we have
\begin{eqnarray}
  \label{eq:FRG}
  && \d_k\K_n = \frac{2k C_{nn}}{(\bar E + k^2)^2},\nn
  && \d_k\eta_n(p) = \frac{2k}{(\bar E + k^2)^2}
     \sum_{m\neq n} \frac{\eta_m(p) C_{mn}}{\K_n-\K_m}.
\end{eqnarray}

The numerical solution of this system allows us to assess the necessary number of bound state fields to achieve satisfactory accuracy for the lower energy eigenvalues. We use discrete momentum values
\begin{equation}
  p_\ell = \left(\ell+\frac12\right) dp,\qquad \ell\in\{0,1,2,\dots N-1\}.
\end{equation}
From the function $\eta_n(p_\ell)$ we create a matrix
\begin{equation}
  \eta_{\ell n}= \sqrt{\frac{dp}{2\pi^2}}\,\eta_n(p_\ell)
\end{equation}
This matrix is normalized to be unitary, since
\begin{equation}
  \delta_{nm} = \frac1{2\pi^2} \int\limits_0^\infty\!dp\,
  \eta_m^*(p)\eta_n(p) \to \frac{dp}{2\pi^2} \sum_{\ell=0}^{N-1}
  \sqrt{\frac{2\pi^2}{dp}} \eta^*_{\ell m} \sqrt{\frac{2\pi^2}{dp}}
  \eta_{\ell n} = \sum_{\ell=0}^{N-1} \eta^\dagger_{m\ell}\eta_{\ell n}.
\end{equation}
We can treat the collection of eigenvalues $\K_n\to\bm\K$ as a vector.

In order to compute $C_{nm}$ from \eqref{eq:etaCdefs} we can introduce a projector $P_{q\ell} = \Theta(k-q) \delta_{q\ell}$, then $\bm C=\bm \eta^\dagger \bm P\bm \eta$. Technically, to avoid adding a lot of zeros, it is simpler to introduce a non-square matrix $\bar\eta_{\ell n} = \eta_{\ell n}$, for $\ell<\kappa$, then $\bm C = \bm{\bar \eta}^\dagger\bm{\bar\eta}$. In addition, we still need an antihermitean matrix $D_{m\neq n} = \frac{C_{mn}}{\K_n-\K_m}$, with $D_{nn}=0$.

Then we pursue the following algorithm.
\begin{enumerate}
\item {\bf Parameter setting}: the external parameters are the value
  of the energy $\bar E$, the number of the points $N$ and the
  resolution of the integrals $dp$. We need $\Lambda=Ndp >\bar E$.
\item {\bf Initialization}: we diagonalize the potential: in matrix
  notation $\bm V$ is a symmetric (in general hermitean) matrix, and
  we are looking for $\bm\eta$ orthogonal (in general unitary) matrix that
  satisfies
  \begin{equation}
    \bm\eta^T \bm V \bm\eta = \mathrm{diag}(V_n).
  \end{equation}
  The starting value for the scale $k=\Lambda=Ndp$, and the initial
  value for $\bm\K$ reads
  \begin{equation}
\label{eq:kinit}
    \K_n = \frac1{V_n}-\frac1{\bar E + \Lambda^2}.
  \end{equation}
\item {\bf Recursion}: for a given $k=\kappa dp$ (where
  $\kappa=N,\dots ,1$) we determine the $\kappa\times N$ matrix
  $\bm{\bar\eta}$, and from that we determine the symmetric $\bm C$
  and antisymmetric $\bm D$ matrices as
  \begin{equation}
    \bm C=\bm{\bar \eta}^T\bm{\bar\eta},\qquad D_{nm} =
    \frac{C_{nm}}{\K_m-\K_n}.
  \end{equation}
  Then we update
  \begin{equation}
    \K_n \to \K_n - \alpha dp\, C_{nn},\qquad 
    \bm\eta \to \bm\eta (\bm1- \alpha dp\,\bm D),\qquad
    \alpha=\frac{2k}{(\bar E+k^2)^2}.
  \end{equation}
\item {\bf End}: do the recursion until $\kappa=1$. Performing the
  above algorithm for different $\bar E$ values, we obtain the function $\K_n(\bar E)$
 which is just the  bound state kernel. The location of the crossing $\K_n(\bar E_n)=0$ provides the bound state
  energies $\bar E_n$.
\end{enumerate}

\section{Charge conjugation}
\label{sec:App}

Although it is a standard piece of knowledge, for completeness we describe charge conjugation operations. Let us start from a Dirac-like Lagrangian
\begin{equation}
  {\cal L} = \bar\zeta (i\gamma^\mu\d_\mu - e\gamma^\mu A_\mu -m)\zeta,
\end{equation}
where $\zeta$ is an anticommuting field. Choose new degrees of freedom as $\zeta = C \chi^*$
\begin{eqnarray}
  {\cal L} 
  &&= \chi^T C^\dagger \gamma_0 (i\gamma^\mu\d_\mu
  -e\gamma^\mu A_\mu -m) C\chi^* = -\bar \chi \gamma_0 C^T
  (-i\gamma^{\mu T} \d_\mu -e\gamma^{\mu T} A_\mu -m)\gamma_0 C^* \chi
  =\nn
  && = \bar \chi \gamma_0 C^T (i\gamma^{\mu T} \d_\mu + e\gamma^{\mu T}
  A_\mu +m)\gamma_0 C^* \chi,
\end{eqnarray}
where we have performed a partial integration, used the anticommuting nature of $\chi$, and used the fact that $\gamma_0^T=\gamma_0$. Now we require
\begin{equation}
  \gamma_0 C^T \gamma^{\mu T} \gamma_0 C^* = \gamma^\mu,\qquad
  \gamma_0 C^T \gamma_0 C^* = -1,
\end{equation}
then we obtain
\begin{equation}
  {\cal L} = \bar \chi(i\gamma^\mu \d_\mu + e\gamma^\mu A_\mu - m)\chi,
\end{equation}
the same form with opposite charge. We use that $\{\gamma_\mu,\gamma_\nu\} = 2g_{\mu\nu}$, that in the Dirac or Weyl representation $\gamma_2\gamma_\mu^T\gamma_2= \gamma^\mu$, and $\gamma_0\gamma_\mu\gamma_0= \gamma^\mu$. Both in Dirac and Weyl representation $\gamma_2^T=\gamma_2$ and $\gamma_2^*=-\gamma_2$. Thus a good choice is $C=\alpha\gamma_2$ where $|\alpha|^2=1$. In this
case, namely
\begin{equation}
  \gamma_0 \alpha \gamma_2^T \gamma^{\mu T} \gamma_0 \alpha^*\gamma_2^* =
  -\gamma_0 \gamma_2 \gamma^{\mu T} \gamma_0 \gamma_2 =
  \gamma_0 \gamma_2 \gamma^{\mu T} \gamma_2 \gamma_0 =
  \gamma_0\gamma_\mu \gamma_0 = \gamma^\mu,
\end{equation}
and
\begin{equation}
  \gamma_0 \alpha \gamma_2^T \gamma_0 \alpha^*\gamma_2^* = 
  -\gamma_0 \gamma_2\gamma_0 \gamma_2 = 
  \gamma_0 \gamma_0\gamma_2 \gamma_2 = -1.
\end{equation}

If we take two fields $\zeta$ and $\psi$ with the same charge, then a mass term of a form $\bar\psi\zeta+\bar\zeta\psi$ is possible. In the language of the $\chi$ fields we have
\begin{equation}
  \bar\zeta\psi = \chi^T C^\dagger \gamma_0 \psi,\qquad
  \bar\psi\zeta = \bar\psi C \gamma_0 (\bar\chi)^T.
\end{equation}
Choosing $C^\dagger=C$ means that a relativistic invariant term of the form
\begin{equation}
  \chi^T {\cal C} \psi + \bar\psi\, {\cal C} (\bar\chi)^T
\end{equation}
is allowed. The hermiticity of $C$ can be ensured by $\alpha=-i$, then
\begin{equation}
  {\cal C} = i\gamma_0\gamma_2.
\end{equation}

If we want to work with the normal adjoint fields instead, then the Lorentz invariant forms are
\begin{equation}
  \psi^\dagger\gamma_0\zeta+\zeta^\dagger\gamma_0\psi = \alpha
  \psi^\dagger \gamma_0 \gamma_2\chi^* + \alpha^* \chi^T
  \gamma_2^\dagger \gamma_0\psi.
\end{equation}
Then it is advantageous to choose $\alpha=1$ and ${\cal C}_E=\gamma_0\gamma_2$, then
\begin{equation}
  \psi^\dagger\gamma_0\zeta+\zeta^\dagger\gamma_0\psi =
  \psi^\dagger{\cal C}_E \chi^* + \chi^T{\cal C}_E\psi.
\end{equation}
In this case ${\cal C}_E^\dagger = {\cal C}_E$.

Similarly, we can consider the vector operators $\bar\psi \Gamma^{Rs} \zeta$ which transforms under the Lorentz group as the $R$ irreducible representation with a definite parity. Here $R=\{S,P,V,A,T\}$ are the scalar, pseudoscalar, vector, axialvector and tensor representations, the corresponding $\Gamma^{Rs}$ matrices are $\Gamma^{Rs} = \{1, \gamma_5, \gamma_\mu, \gamma_5\gamma_\mu,\sigma_{\mu\nu}\}$, respectively. We can rewrite them with the $\chi$ field as
\begin{equation}
  \psi^\dagger \gamma_0 \Gamma^{Rs} \zeta = \psi^\dagger
  \bar\Gamma^{Rs}\gamma_0 C\gamma_0 \chi^* =  \psi^\dagger
  \bar\Gamma^{Rs} {\cal C}_E \chi^*,
\end{equation}
where $\bar\Gamma^{Rs}=\gamma_0\Gamma^{Rs}\gamma_0$.

\end{document}